\documentclass[preprint]{aastex}

\newcommand{\be}{\begin{equation}}
\newcommand{\ee}{\end{equation}}
\newcommand{\bfv}{{\bf v}}
\newcommand{\bfr}{{\bf r}}
\newcommand{\bnabla}{\mbox{\boldmath $\nabla$}}

\newcommand{\pd}{\partial }
\newcommand{\half}{{\textstyle{1\over2}}}

\begin{document}

\title{Slow modes in Keplerian disks}

\author{Scott Tremaine}

\affil{Princeton University Observatory, Peyton Hall, \\
Princeton, NJ 08544-1001}
\email{tremaine@astro.princeton.edu}

\begin{abstract}

\noindent
Low-mass disks orbiting a massive body can support ``slow'' normal modes, in
which the eigenfrequency is much less than the orbital frequency. Slow modes
are lopsided, i.e., the azimuthal wavenumber $m=1$. We investigate the
properties of slow modes, using softened self-gravity as a simple model for
collective effects in the disk. We employ both the WKB approximation and
numerical solutions of the linear eigenvalue equation. We find that all slow
modes are stable. Discrete slow modes can be divided into two types, which we
label g-modes and p-modes. The g-modes involve long leading and long trailing
waves, have properties determined by the self-gravity of the disk, and are
only present in narrow rings or in disks where the precession rate is
dominated by an external potential. In contrast, the properties of p-modes are
determined by the interplay of self-gravity and other collective
effects. P-modes involve both long and short waves, and in the WKB
approximation appear in degenerate leading/trailing pairs. Disks support a
finite number---sometimes zero---of discrete slow modes, and a continuum of
singular modes.

\end{abstract}


\section{Introduction}

\noindent
Disks orbiting massive bodies are found in many astrophysical systems. These
may be composed of gas (surrounding protostars, compact objects in binary star
systems, or black holes in galactic nuclei), dust (a component of
protoplanetary disks), solid bodies (planetary rings and planetesimal disks),
or stars (galactic nuclei).  Normally the disk mass $M_d$ in these systems is
much less than the central mass $M$, so that the gravitational potential is
nearly Keplerian. A special feature of the Keplerian potential is that
eccentric orbits do not precess. This feature arises from a 1:1 resonance
between the radial and azimuthal frequencies, and implies that the evolution
of the eccentricity distribution in an isolated Keplerian disk is determined
by the collective effects in the disk (self-gravity, pressure, collisions,
viscosity, etc.), no matter how small these may be. Thus we expect that
Keplerian disks may be capable of supporting a set of normal modes with
eigenfrequency $\omega$ much less than the Keplerian orbital frequency
$\Omega(r)=(GM/r^3)^{1/2}$, and properties determined by weak collective
effects in the disk. In the linear limit these modes have azimuthal wavenumber
$m=1$. We shall call them ``slow'' modes.

The goal of this paper is to investigate slow modes in Keplerian disks.  In
this Section we provide an introduction and orientation based on the WKB
approximation. We shall find that in many cases slow modes have scales
comparable to the size of the disk, so that the WKB analysis must be
supplemented by numerical calculations of large-scale modes. We soften the
self-gravity of the disk to mimic the effects of pressure in fluid disks and
velocity dispersion in collisionless disks, and derive the integral eigenvalue
equation that describes large-scale slow modes in \S II. We discuss the
numerical solution of this integral equation for several disk models in \S
III. Section IV contains a discussion of earlier related work, and \S V
contains conclusions.

\subsection{The precession rate}

\noindent
We employ cylindrical coordinates $(r,\phi,z)$ with origin at the central mass
$M$, and approximate the unperturbed disk as razor-thin, with surface density
$\Sigma_d(r)$ in the $z=0$ plane. We assume that the disk mass is small,
$M_d/M\sim\Sigma_dr^2/M\sim \epsilon\ll1$. The unperturbed disk material
orbits in the axisymmetric potential
\be
\Phi(r)=-{GM\over r}+\Phi_d(r)+\Phi_e(r),
\ee
where $\Phi_d(r)$ is the potential arising from the self-gravity of the disk
and $\Phi_e(r)$ is the non-Keplerian potential from any external source, and
both are assumed to be O$(\epsilon)$. 

For nearly circular orbits, the azimuthal and radial frequencies $\Omega>0$
and $\kappa>0$ are given by (e.g. \nocite{bt87}Binney and Tremaine 1987)
\begin{eqnarray}
\Omega^2(r) & = & {GM\over r^3}+{1\over r}{d\over dr}(\Phi_d+\Phi_e), 
\nonumber \\
\kappa^2(r) & = & {GM\over r^3}+{d^2\over dr^2}(\Phi_d+\Phi_e)+
{3\over r}{d\over dr}(\Phi_d+\Phi_e). 
\label{eq:kapdef}
\end{eqnarray}
The line of apsides of an eccentric test-particle orbit subjected only to
gravity precesses at a rate
\begin{eqnarray}
\dot\varpi & = & \Omega-\kappa \nonumber \\
           & = &-{1\over 2\Omega(r)}\left({2\over r}{d\over dr}
+{d^2\over dr^2}\right)(\Phi_d+\Phi_e)+\hbox{O}(\epsilon^2),
\label{eq:precdef}
\end{eqnarray}
where from now on $\Omega(r)$ is assigned the Keplerian value,
$(GM/r^3)^{1/2}$.

\subsection{The Kuzmin disk}

\label{sec:kuzmin}

\noindent
Several of our examples will be based on the Kuzmin disk, which has surface
density and potential (e.g. \nocite{bt87}Binney and Tremaine 1987)
\be
\Sigma_d(r)={aM_d\over2\pi(r^2+a^2)^{3/2}}, \qquad \Phi_d(r)=-{GM_d\over
(r^2+a^2)^{1/2}},
\label{eq:kuzmin}
\ee
where $M_d$ is the disk mass and $a$ is the disk scale length. Using equation
(\ref{eq:precdef}), the precession rate due to the Kuzmin disk is 
\be
\dot\varpi_d=-{3GM_da^2\over 2\Omega(r)(r^2+a^2)^{5/2}},
\ee
for $M_d\ll M$.  The precession rate is negative at all radii,
approaches zero as $r^{3/2}$ for small radii and as $r^{-7/2}$ at large radii,
and has a single extremum at $r_0=(\frac{3}{7})^{1/2}a=0.6547a$ where
$\dot\varpi(r_0)=-0.3257(M_d/M)(GM/a^3)^{1/2}$.

\subsection{The WKB approximation}

\label{sec:wkb}

\noindent
We can write the linearized surface-density response of the disk as a
superposition of terms of the form
$\Sigma(r,\phi,t)=A(r)\exp\left\{i[\int^rk(r)dr+m\phi-\omega t]\right\}$,
where $m$ is the azimuthal wavenumber, $k(r)$ is the radial wavenumber, and
$\omega$ is the frequency. In the WKB or tight-winding approximation where
$|kr|\gg 1$, the dispersion relation between $k(r)$ and $\omega$ depends only
on the local properties of the disk, and can be derived analytically for both
fluid and collisionless disks. 

For a barotropic fluid disk with sound speed $c(r)$, the WKB dispersion
relation reads (Safronov 1960; Binney and Tremaine 1987)\nocite{saf60,bt87}
\be
(\omega-m\Omega)^2=\kappa^2-2\pi G\Sigma_d|k|+k^2c^2.
\label{eq:disprel}
\ee
This relation implies that the disk is locally stable to axisymmetric ($m=0$)
disturbances if and only if
\be
Q\equiv {c\kappa\over \pi G\Sigma_d} > 1.
\label{eq:qgas}
\ee
For thin, low-mass disks we have $c\ll\Omega r$ and $\Sigma_d\ll M/r^2$, so
that in general the dispersion relation (\ref{eq:disprel}) can only be
satisfied if $|kr|\gg 1$, which justifies the use of the WKB
approximation. However, the special case of $m=1$ disturbances in a nearly
Keplerian disk is different. Using equation (\ref{eq:precdef}) the dispersion
relation can be rewritten in the form
\be
\omega=\dot\varpi +{\pi G\Sigma_d|k|\over\Omega} - {k^2c^2\over 2\Omega}
+{1\over\Omega}\hbox{O}(\dot\varpi^2,\omega^2).
\label{eq:dispone}
\ee
Since we are interested in slow modes in nearly Keplerian disks,
the terms of order $\dot\varpi^2$, $\omega^2$ can be dropped. At a given
radius, the maximum value of $\omega$ is attained at wavenumber
$|k|\equiv k_0=\pi G\Sigma_d/c^2$. 

Let us neglect the effects of pressure for the moment, by dropping the term
proportional to $c^2$ in equation (\ref{eq:dispone}).  Thus the WKB dispersion
relation for slow waves dominated by self-gravity becomes 
\be
\omega=\dot\varpi +{\pi G\Sigma_d|k|\over\Omega}. 
\label{eq:wkb}
\ee
Since $\dot\varpi/\Omega=\hbox{O}(\Sigma_dr^2/M)$ for isolated disks, in
general the two terms on the right side of (\ref{eq:dispone}) are only
comparable if $|kr|\sim 1$; thus slow gravity-dominated waves must be
large-scale (i.e. the WKB approximation is {\em not} satisfied) and have
eigenfrequency $\omega=\hbox{O}(\dot\varpi)$. The neglect of the pressure term
is valid for such waves so long as $G\Sigma_dr\gg c^2$, or ${\cal M}\gg Q$,
where ${\cal M}(r)\equiv\Omega r/c$ is the Mach number.

Next consider a collisionless disk with a Schwarzschild distribution function
having radial velocity dispersion $c_R(r)$. The WKB dispersion relation reads
(Kalnajs 1965; Lin and Shu 1966; Binney and Tremaine
1987)\nocite{kal65,ls66,bt87} 
\be
(\omega-m\Omega)^2=\kappa^2-2\pi G\Sigma_d|k|{\cal
F}\left({\omega-m\Omega\over\kappa},{k^2c_R^2\over\kappa^2}\right),
\label{eq:disprels}
\ee
where ${\cal F}(s,\chi)$ is the reduction factor. Explicit formulae for the
reduction factor are given in the references above. The disk is locally stable
to axisymmetric disturbances if and only if (Toomre 1964)\nocite{too64}
\be
Q\equiv {c_R\kappa\over 3.36 G\Sigma_d} > 1,
\ee
which closely resembles equation (\ref{eq:qgas}).  For $m=1$ disturbances in a
nearly Keplerian disk, the dispersion relation (\ref{eq:disprels}) simplifies
to
\be
\omega=\dot\varpi +{\pi G\Sigma_d|k|\over\Omega}{\cal F}_K
\left({k^2c_R^2\over\Omega^2}\right),
\label{eq:dispxxx}
\ee
in which
\be
{\cal F}_K(\chi)={2\over\chi}e^{-\chi}I_1(\chi),
\ee
and $I_1(\chi)$ is a modified Bessel function. At a given radius, the
maximum value of $\omega$ is attained at wavenumber $|k|\equiv
k_0=0.7643\,\Omega/c_R$. Since both terms on the right side of
(\ref{eq:dispxxx}) scale as the disk surface density $\Sigma_d$, for a fixed
velocity-dispersion profile the shape of a mode is independent of the
disk mass and the eigenfrequency is proportional to the disk mass.

For the moment assume $\chi\ll1$; then the reduction factor ${\cal F}(\chi)=1$
and the dispersion relation (\ref{eq:dispxxx}) reduces to (\ref{eq:wkb}): we
conclude, not surprisingly, that the WKB dispersion relation for slow gravity
waves is the same for fluid and collisionless disks. The neglect of the
reduction factor is valid for such waves so long as $\chi\ll1$ when $|kr|\sim
1$, which in turn requires that the Mach number ${\cal M}_R\equiv \Omega
r/c_R\gg 1$.

Many of the properties of waves in collisionless disks can be mimicked by
softening the gravitational potential, using $-Gm/(d^2+b^2)^{1/2}$ instead of
$-Gm/d$ for the potential due to a mass $m$ at distance $d$. Here the
parameter $b$ is the softening length. This modification yields the WKB
dispersion relation (Miller 1971; Erickson 1974; Toomre
1977)\nocite{mil71,eri74,too77} 
\be
\omega=\dot\varpi(r) +{\pi G\Sigma_d(r)\over\Omega(r)}|k|\exp(-|k|b).
\label{eq:dispsoft}
\ee 
At a given radius, the maximum value of $\omega$ is attained at wavenumber
$|k|\equiv k_0=1/b$. As in the case of a collisionless disk, both terms on the
right side scale as the disk surface density $\Sigma_d$, so the eigenfrequency
is proportional to the disk mass. In the long-wavelength limit, $|k|b\ll1$,
the dispersion relation simplifies once again to equation (\ref{eq:wkb}).

By analogy with the theory of WKB waves in galactic disks,
we call waves trailing if $k>0$ and leading if $k<0$. Long waves have
$|k|<k_0$ and short waves have $|k|>k_0$. 

We conclude from these approximate arguments that a low-mass, thin disk (by
which we mean $\Sigma_dr^2/M\sim\epsilon\ll1$, and ${\cal M}\gg Q$ for a fluid
disk, or ${\cal M}\gg 1$ for a collisionless disk) could support slow modes
with the following properties: (i) the azimuthal wavenumber $m=1$; (ii) the
dominant collective effect is self-gravity; (iii) $|kr|=\hbox{O}(1)$, that is,
the wavelength is comparable to the disk radius, the properties of the mode
are determined globally rather than locally, and the mode cannot be described
accurately with the WKB approximation; (iv) the properties of the mode depend
on the disk self-gravity but not on other collective effects; (v) the
eigenfrequency is of order $\epsilon\Omega$. Many of these conclusions were
derived already by \nocite{lg99}Lee and Goodman (1999), in a paper focused
mainly on nonlinear waves in nearly Keplerian disks.

\begin{figure}
\plotone{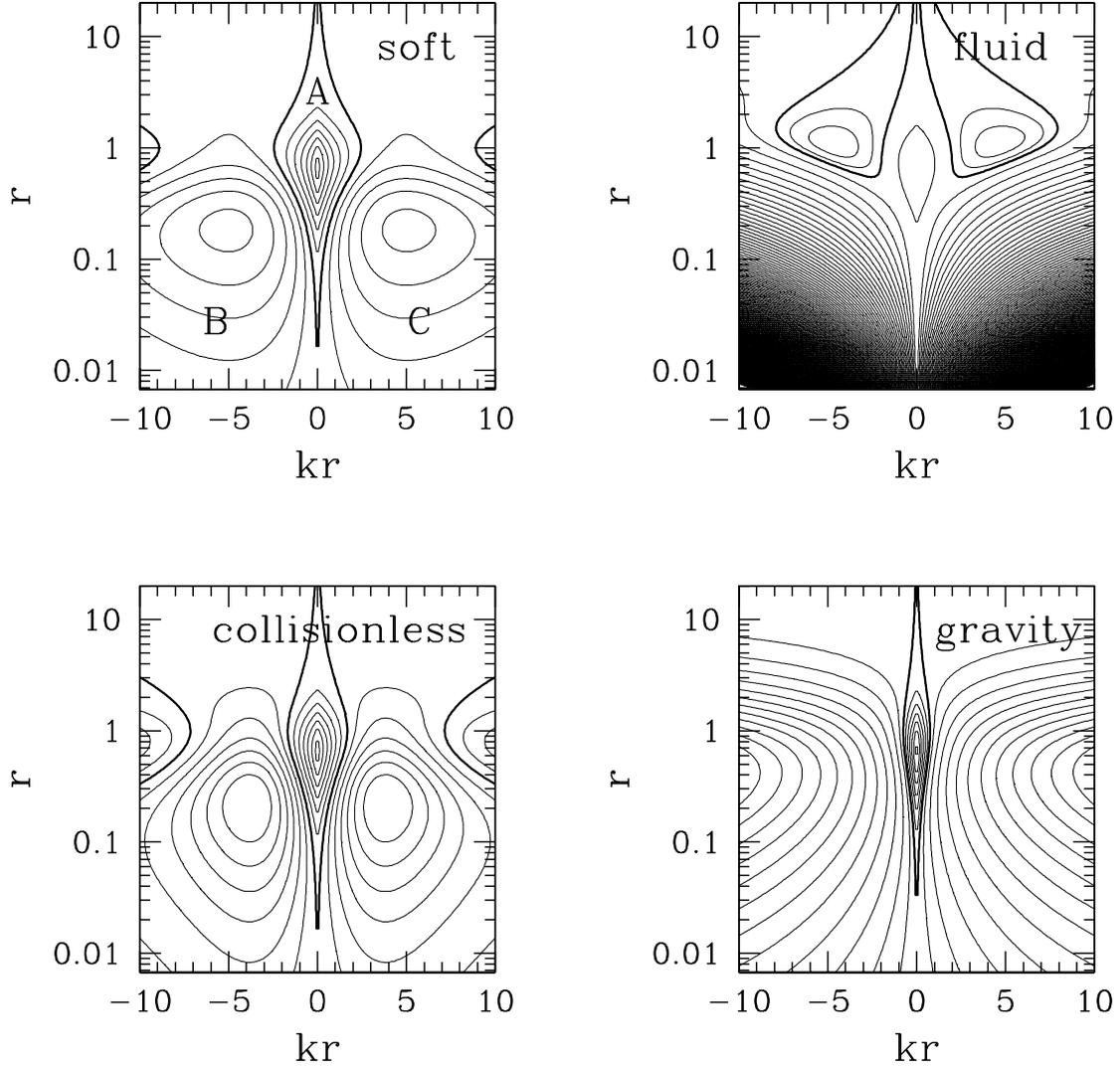}
\caption{Contours of constant frequency $\omega$ for WKB slow waves in a
Kuzmin disk (\S \ref{sec:kuzmin}), as a function of wavenumber $kr$ and radius
$r$. The dispersion relations shown are for a disk with softened gravity
(eq. \ref{eq:dispsoft}), a fluid disk (eq. \ref{eq:dispone}), a collisionless
disk (eq. \ref{eq:dispxxx}), and a gravity-dominated disk (eq. \ref{eq:wkb}).
The disk scalelength $a=1$, the softening length $b=0.2r$, the Mach number
${\cal M}=\Omega /cr$ or ${\cal M}_R=\Omega r/c_R$ is the same for the fluid
and the collisionless disk and equal to 5 at all radii.  Both the
gravitational constant $G$ and the central mass $M$ are unity; the disk mass
$M_d=1$ and $\omega$ scales as $M_d$ so long as $b$, ${\cal M}_R$ or $M_d{\cal
M}^2$ remain invariant. The extrema at $kr=0$ are minima. The contours in the
right-hand panels are not uniformly spaced.}
\label{fig:wkb}
\end{figure}

Despite conclusion (iii), the WKB approximation provides an invaluable
guide to the properties of slow modes.  It is useful to plot contours of
constant $\omega$ in the $(k,r)$ plane. Figure \ref{fig:wkb} shows such
contour plots for the Kuzmin disk discussed in \S \ref{sec:kuzmin} and four
dispersion relations for slow waves: a disk with softened gravity
(eq.~\ref{eq:dispsoft}), a fluid disk (eq.~\ref{eq:dispone}), a collisionless
disk (eq.~\ref{eq:dispxxx}), and a gravity-dominated disk
(eq.~\ref{eq:wkb}). Local maxima in the contour plots occur at $k=\pm k_0$,
and a local minimum at $k=0$.

In these diagrams, a WKB wavepacket travels along a contour of constant
$\omega$, at a radial speed given by the group velocity ${d\omega/ dk}$
\nocite{too69}(Toomre 1969). Long trailing and short leading waves propagate
outwards, while long leading and short trailing waves propagate in.  In the
WKB approximation a mode corresponds to a closed contour in Figure
\ref{fig:wkb} on which the total phase change $\oint k(r)dr$ over one circuit
is an integer multiple of $2\pi$, including possible phase shifts at the
turning points. The models in the Figure exhibit two types of closed contour:

\begin{enumerate}

\item On contours such as C in the upper-left panel, an
outward-traveling long trailing wave refracts into an inward-traveling short
trailing wave, which in turn refracts back into a long wave. Contours such as
B are similar, except that they involve inward-traveling long leading waves
and outward-traveling short leading waves. WKB modes of types B and C
occur in degenerate pairs. Modes of this kind are not present in the
gravity-dominated disk (lower right panel), since this disk supports only long
waves. In the Kuzmin disk these modes have $\omega>0$, and the modes with the
smallest phase change around the contour have the highest frequency and
wavenumbers localized near $\pm k_0$. We call these {\em p-modes}, since they
require the presence of pressure, velocity dispersion, or softening.

\item In contrast, closed contours such as A involve only long waves, and
involve reflection of an outward-traveling long trailing wave into an
inward-traveling long leading wave, which then reflects back to long
trailing. These modes are present in all four panels; their properties are
similar in all panels since they depend only on the surface-density
distribution in the disk and not other collective effects. In the Kuzmin disk
these modes have $\omega<0$, and the modes with the fewest nodes have the most
negative values of $\omega$.  The WKB approximation is highly suspect for
waves of this kind since $|kr|\lesssim 1$, so numerical mode calculations are
required. We call these {\em g-modes}, since self-gravity is the only
collective effect involved.

\end{enumerate}

Figure \ref{fig:wkb} suggests that the WKB dispersion relation for
collisionless Keplerian disks can be reproduced quite well by disks with
softened gravity \nocite{eri74,too77}(Erickson 1974; Toomre 1977); and at least
the qualitative features of the dispersion relation in fluid disks are similar
as well. Therefore, in our numerical mode calculations we shall soften the
self-gravity of the disk but neglect other collective effects.

\section{Linear perturbation theory}

\label{sec:linear}

\noindent
In this Section we derive the linear eigenvalue equation for slow modes in
disks with softened gravity.  We write the surface density, velocity and
potential of the perturbed disk as $\Sigma_d(r)+\Sigma_a(\bfr,t)$,
$\bfv_d(\bfr)+\bfv_a(\bfr,t)=r\Omega\hat{\mbox{\boldmath
$\phi$}}+\bfv_a(\bfr,t)= r\Omega\hat{\mbox{\boldmath
$\phi$}}+u_a(\bfr,t)\hat\bfr+v_a(\bfr,t)\hat{\mbox{\boldmath
$\phi$}}$, $\Phi_d(r)+\Phi_a(\bfr,t)$.  The linearized Euler and continuity
equations read
\begin{eqnarray}
{\pd\bfv_a\over\pd t}+(\bfv_d\cdot\bnabla)\bfv_a+(\bfv_a\cdot\bnabla)\bfv_d & =
& -\bnabla\Phi_a, \nonumber \\
{\pd \Sigma_a\over\pd t}+\bnabla\cdot(\Sigma_d\bfv_a+\Sigma_a\bfv_d) & = & 0.
\label{eq:ggg}
\end{eqnarray}
We write the perturbation variables $(u_a,v_a,\Sigma_a,\Phi_a)$ in the form
$X_a(r,\phi,t)=X_a^m(r)\exp[i(m\phi-\omega t)]$. Equations (\ref{eq:ggg})
become
\begin{eqnarray}
i(m\Omega-\omega)u_a^m-2\Omega v_a^m & = & -{d\Phi_a^m\over dr}, \nonumber \\
{\kappa^2\over2\Omega}u_a^m+i(m\Omega-\omega)v_a^m & = & 
-{im\over r}\Phi_a^m,\nonumber\\
i(m\Omega-\omega)\Sigma_a^m & = & -{1\over r}{d\over dr}(r\Sigma_du_a^m)
- {im\over r}\Sigma_dv_a^m.
\label{eq:vdef}
\end{eqnarray}
The first two of these can be solved to yield
\begin{eqnarray}
u_a^m & = & -{i\over D}\left[(m\Omega-\omega){d\over dr}+{2m\Omega\over
r}\right]\Phi_a^m, \nonumber \\
v_a^m & = & {1\over D}\left[{\kappa^2\over2\Omega}{d\over dr}+{m\over
r}(m\Omega-\omega)\right]\Phi_a^m,
\label{eq:uvdef}
\end{eqnarray}
where 
\be
D=\kappa^2-(m\Omega-\omega)^2.
\label{eq:dendef}
\ee 

\subsection{Nearly Keplerian disks}

\noindent
We now specialize to slow perturbations in nearly Keplerian disks. Thus we
assume that the disk potential and any external non-Keplerian potential
satisfy $|\Phi_d,\Phi_e|/(GM/r)=\hbox{O}(\epsilon)\ll 1$, and that the
azimuthal wavenumber $m=1$. Furthermore we assume
$|\omega|/\Omega=\hbox{O}(\epsilon)$, for the reasons given in \S
\ref{sec:wkb}. Using equation (\ref{eq:precdef}), the denominator $D$
(eq. \ref{eq:dendef}) becomes
\be
D=2\Omega(\omega-\dot\varpi)+\hbox{O}(\epsilon^2);
\ee
and equations (\ref{eq:uvdef}) become
\begin{eqnarray}
u_a^1 & = & -{i\over 2(\omega-\dot\varpi)}\left({d\Phi_a^1\over dr}+
{2\Phi_a^1\over r}\right)
\nonumber \\
v_a^1 & = & {1\over 4(\omega-\dot\varpi)}\left({d\Phi_a^1\over dr}+
{2\Phi_a^1\over r}\right),
\label{eq:vdefa}
\end{eqnarray}
with fractional error O($\epsilon)$. Thus to leading order in $\epsilon$,
\be
u_a^1=-2iv_a^1,
\label{eq:nrkep}
\ee
which is simply the relation between the radial and azimuthal non-circular
velocities for a nearly circular Keplerian orbit. The Keplerian eccentricity
$e_K$ is related to the perturbed velocities by
\be
u_a^1=-2iv_a^1=-i\left(GM\over r\right)^{1/2}e_K\exp(-i\varpi).
\label{eq:eccdef}
\ee

We may use equation (\ref{eq:nrkep}) to eliminate the radial velocity from the
continuity equation (the last of eqs. 
\ref{eq:vdef}); to leading order in $\epsilon$ this now reads 
\be
\Omega\Sigma_a^1={2\over r}{d\over dr}(r\Sigma_dv_a^1)-{\Sigma_dv_a^1\over r}=
2{d\over dr}(\Sigma_dv_a^1)+{\Sigma_dv_a^1\over r}.
\label{eq:sigdef}
\ee

The second of equations (\ref{eq:vdefa}) can be rewritten as
\be
(\omega-\dot\varpi)v_a^1={1\over 4}{d\Phi_a^1\over dr}+{\Phi_a^1 \over 2r}.
\label{eq:vvdef}
\ee
Equations (\ref{eq:sigdef}) and (\ref{eq:vvdef}), together with Poisson's
equation relating $\Phi_a^1$ and $\Sigma_a^1$, describe the slow modes of
Keplerian disks when self-gravity is the only collective effect.

\subsection{Potential theory}

\label{sec:pot}

\noindent
Poisson's equation can be written as
\be
\Phi_a^m(r)=\int_0^\infty dr'r'P_m(r,r')\Sigma_a^m(r');
\label{eq:poisdef}
\ee
the same relation, with $m=0$, also holds between the potential $\Phi_d$ and
the surface density $\Sigma_d$ of the unperturbed disk.  The kernel in
equation (\ref{eq:poisdef}) is
\be
P_m(r,r')= -{\pi G\over r_>}b_{1/2}^{(m)}(r_</r_>)+{\pi G r\over
{r'}^2}(\delta_{m1}+\delta_{m,-1}).
\label{eq:poiss}
\ee
Here $r_<=\hbox{min\,}(r,r')$, $r_>=\hbox{max}(r,r')$, and the second
term is the indirect potential arising from the motion of the central body in
response to the disk perturbation. The Laplace coefficient 
\be
b_{1/2}^{(m)}(\alpha)={2\over\pi}\int_0^\pi {\cos m\theta\,d\theta\over
(1-2\alpha\cos\theta +\alpha^2)^{1/2}}
\ee
\nocite{mur99}(Murray and Dermott 1999). In particular,
\be
b_{1/2}^{(0)}(\alpha)={4\over \pi}K(\alpha), \qquad b_{1/2}^{(1)}(\alpha)
={4\over\pi \alpha}[K(\alpha)-E(\alpha)],
\label{eq:bdef}
\ee
where $K(\alpha)$ and $E(\alpha)$ are complete elliptic integrals. 
We shall use the relations
\be
{dE(\alpha)\over d\alpha}={E(\alpha)-K(\alpha)\over\alpha}, \qquad
{dK(\alpha)\over
d\alpha}={E(\alpha)\over\alpha(1-\alpha^2)}-{K(\alpha)\over\alpha}.
\label{eq:ellder}
\ee

The Laplace coefficients are singular at $\alpha=1$, 
\be
b_{1/2}^{(0)}(1-\delta)=-{2\over\pi}\ln\delta+{2\over\pi}\ln
8+\hbox{O}(\delta\ln\delta);\quad
b_{1/2}^{(1)}(1-\delta)=-{2\over\pi}\ln\delta+{2\over\pi}(\ln
8-2)+\hbox{O}(\delta\ln\delta).
\label{eq:lapsing}
\ee

We can soften the logarithmic singularity by replacing $\ln\delta$ by
$\half\ln(\delta^2+\beta^2)$ where $\beta$ is the softening parameter; the
actual softening length is then radius-dependent, $b=\beta r$. If we adopt
this prescription then we must soften the stronger singularities that arise
from derivatives of $\ln\delta$ by differentiating
$\half\ln(\delta^2+\beta^2)$. Thus
\be
\ln\delta\to \half\ln(\delta^2+\beta^2), \qquad {1\over\delta}\to {\delta\over 
\delta^2+\beta^2}, \qquad {1\over\delta^2}\to
{\delta^2-\beta^2\over(\delta^2+\beta^2)^2}.
\label{eq:soften}
\ee
A full description of the softening procedure follows equations
(\ref{eq:narrow}) and (\ref{eq:wtq}).

\subsection{The eigenvalue equation}

\label{sec:eigen}

\noindent
Combining equations (\ref{eq:sigdef}), (\ref{eq:vvdef}) and
(\ref{eq:poisdef}), we get
\be
[\omega-\dot\varpi(r)]v_a^1(r)=\int {r'dr'\over\Omega(r')}\left[{1\over
4}{\partial\over\partial r}P_1(r,r')+{1\over 2r}P_1(r,r')\right]\left[2{d\over
dr'}\Sigma_d(r')v_a^1(r') +{\Sigma_d(r')v_a^1(r')\over r'}\right].
\ee
This can be recast by writing
\be
y(r)=\left[r^2\Sigma_d(r)\over\Omega(r)\right]^{1/2}v_a^1(r);
\label{eq:eigen}
\ee
we then integrate by parts, using the fact that $\Omega(r)\propto r^{-3/2}$, to
obtain 
\be
\omega y(r)=\dot\varpi(r)y(r)+\int {dr'\over r'}K(r,r')y(r'),
\label{eq:intsymm}
\ee
where the kernel  
\be
K(r,r') =  -2\left[\Sigma_d(r)\Sigma_d(r')\over 
\Omega(r)\Omega(r')\right]^{1/2}\left(1+{1\over
2}{\partial\over\partial\ln r}\right)\left(1+{1\over
2}{\partial\over\partial\ln r'}\right)P_1(r,r').
\ee
The contribution of the indirect potential in equation (\ref{eq:poiss}) to
$K(r,r')$ vanishes, and the remaining component of $P_1(r,r')$ is symmetric in
$r$ and $r'$, so $K(r,r')$ is also symmetric.  

Since the kernel is real and symmetric, the right side of equation
(\ref{eq:intsymm}) can be regarded as a linear, Hermitian operator acting on
$y(r')$. From the properties of Hermitian operators we can conclude (e.g.,
Courant and Hilbert 1953)\nocite{cour} that (i) the eigenvalues $\omega$ are
real; thus {\em all slow disturbances are stable}; (ii) the eigenfunctions
$y_n(r)$ associated with different eigenvalues $\omega_n$ are orthogonal with
respect to the inner product $(y,z)\equiv \int y^\ast(r)z(r) dr/r$, that is,
$(y_n,y_m)=0$ if $\omega_n\not=\omega_m$.

Since the kernel is real, the solutions $y(r)$ or $v_a^1(r)$ can be assumed to
be real. In this case it proves useful to redefine the eccentricity $e_K$
(eq. \ref{eq:eccdef}) by the relation
\be
v_a^1=\half\left(GM\over r\right)^{1/2}e_K(r);
\label{eq:eccdefa}
\ee
so that $e_K(r)$ is real but can have either sign.

Using equations (\ref{eq:poiss}), (\ref{eq:bdef}) and (\ref{eq:ellder}) we
can derive alternative forms for the kernel,
\begin{eqnarray}
K(r,r') &=& -{\pi G\over 2}\left[\Sigma_d(r)\Sigma_d(r')\over rr' 
\Omega(r)\Omega(r')\right]^{1/2}\alpha^{1/2}\left(\alpha^2{d^2\over
d\alpha^2}+ 2\alpha{d\over
d\alpha}-2\right)b_{1/2}^{(1)}(\alpha),\nonumber \\
&=& -G\left[\Sigma_d(r)\Sigma_d(r')\over 
rr'\Omega(r)\Omega(r')\right]^{1/2}C(\alpha),
\end{eqnarray}
where $\alpha=r_</r_>$ and
\be
C(\alpha)=2\alpha^{-1/2}\left[{2(1-\alpha^2+\alpha^4)\over(1-
\alpha^2)^2}E(\alpha)-{2-\alpha^2\over1-\alpha^2}K(\alpha)\right].
\label{eq:cdef}
\ee

The precession rate is given by equations (\ref{eq:precdef}) and 
(\ref{eq:poisdef}) and can be written as 
\be
\dot\varpi=\dot\varpi_d+\dot\varpi_e=\int {dr'\over r'}Q(r,r') + \dot\varpi_e.
\label{eq:precess}
\ee
Here $\dot\varpi_d$ and $\dot\varpi_e$ are the precession rates from the
self-gravity of the disk and from external sources (due to the potentials
$\Phi_d$ and $\Phi_e$ respectively). The kernel is 
\begin{eqnarray}
Q(r,r') & = & {\pi G{r'}^{3/2}\Sigma_d(r')\over 2\Omega(r)r^{5/2}}\alpha^{3/2}
\left(\alpha{d^2\over d\alpha^2}+ 2{d\over d\alpha}\right)
b_{1/2}^{(0)}(\alpha)\nonumber \\
 & = & {G\Sigma_d(r'){r'}^{3/2}\over \Omega(r)r^{5/2}}D(\alpha),
\end{eqnarray}
where
\be
D(\alpha)=2\alpha^{1/2}\left[{1+\alpha^2\over (1-\alpha^2)^2}E(\alpha)
-{1\over 1-\alpha^2}K(\alpha)\right].
\label{eq:ddef}
\ee

For the sake of simplicity we shall simply assume that any external precession
mimics the precession rate due to the disk itself, that is,
$\dot\varpi_e(r)=f\dot\varpi_d(r)$ where $f$ is a constant. In this case the
external precession can be eliminated if we replace the eigenfrequency
$\omega$ by $\omega(1+f)$ and replace the integral eigenvalue equation
(\ref{eq:intsymm}) by
\be
\omega y(r)=\dot\varpi_d(r)y(r)+\lambda\int {dr'\over r'}K(r,r')y(r'),
\label{eq:intsymma}
\ee
where $\lambda=(1+f)^{-1}$. 

\begin{figure}
\plotone{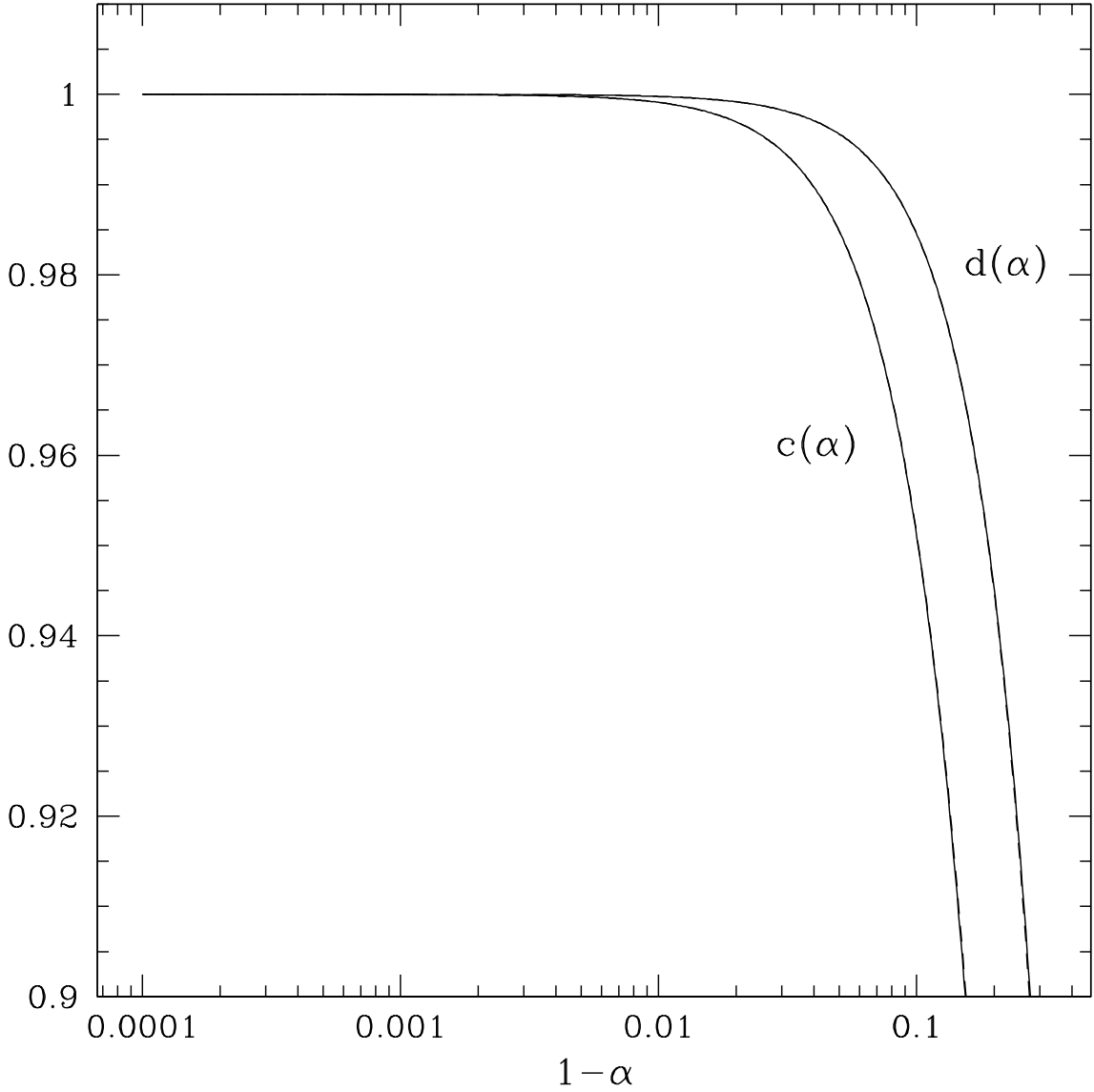}
\caption{The functions $C(\alpha)$ and $D(\alpha)$
(eqs. \ref{eq:cdef} and \ref{eq:ddef}). The dominant singularity has been
removed by plotting $c(\alpha)\equiv x^2C(\alpha)$ and $d(\alpha)\equiv
x^2d(\alpha)$, where $x=-\log\alpha$ (note that $\alpha<1$ so $x>0$). The plot
also contains dashed lines that almost coincide with the solid lines. These
are the approximations used in the numerical integration scheme; the
quantities plotted are $x^2w_K(x)$ (eq. \ref{eq:wt}) and $x^2w_Q(x)$
(eq. \ref{eq:wtq}).}
\label{fig:two}
\end{figure}

The behavior of $C(\alpha)$ and $D(\alpha)$ as $\alpha\to 0$ is given by
\be
C(\alpha)\to {15\pi\over 8}\alpha^{7/2}, \qquad 
D(\alpha)\to {3\pi\over 2}\alpha^{5/2}.
\ee
Both functions are singular at $\alpha=1$: if
we write $x\equiv -\log\alpha>0$ then
\begin{eqnarray} 
C(\alpha) & = & {1\over x^2} + {15\over 8}\log {x\over 8} +
{179\over 48} +\hbox{O}(x),\nonumber \\
D(\alpha) & = & {1\over x^2} + {3\over 8}\log {x\over 8} +
{11\over 48} +\hbox{O}(x).
\label{eq:sing}
\end{eqnarray}
Figure \ref{fig:two} shows the behavior of $C(\alpha)$ and $D(\alpha)$; for
clarity we have removed the dominant singularity by plotting
$c(\alpha)\equiv x^2C(\alpha)$ and $d(\alpha)\equiv x^2d(\alpha)$.

The eigenvalue equation simplifies considerably in the case of narrow rings:
we can replace $C(\alpha)$ and $D(\alpha)$ by the leading term in equations
(\ref{eq:sing}) and need not distinguish $r$ from $r'$ except where they occur 
in rapidly varying functions such as $\Sigma_d(r)$ and
$\dot\varpi(r)$. Equations (\ref{eq:eigen}) and (\ref{eq:precess}) become
\begin{eqnarray}
[\omega-\dot\varpi(r)]v_a^1(r) & = & -{G\over\Omega}\int
{dr'\over (r-r')^2}\Sigma_d(r')v_a^1(r'), \nonumber \\
\dot\varpi(r) & = & {G\over\Omega}\int
{dr'\over(r-r')^2}\Sigma_d(r'),
\label{eq:narrow}
\end{eqnarray}
where $\Omega$ can be treated as constant. The singular kernel can be
softened according to equation (\ref{eq:soften}):
\begin{eqnarray}
[\omega-\dot\varpi(r)]v_a^1(r) & = & -{G\over\Omega}\int
dr'\Sigma_d(r')v_a^1(r'){(r-r')^2-b^2\over [(r-r')^2+b^2]^2}, \nonumber \\
\dot\varpi(r) & = & {G\over\Omega}\int
dr'\Sigma_d(r'){(r-r')^2-b^2\over [(r-r')^2+b^2]^2}.
\label{eq:narrowb}
\end{eqnarray}
A trivial solution of these equations is $\omega=0$, $v_a^1(r)=$constant. 

\subsection{Numerical methods}
 
\noindent
We use similar numerical methods to evaluate the integrals in equations
(\ref{eq:precess}) and (\ref{eq:intsymma}) and discuss both integrals
together. We approximate the integrals using an $N$-point grid that is uniform
in $u\equiv\log r$.  The main complication is the singularity of the
integrands when $r$ is close to $r'$. To handle this singularity, we write the
integral in (\ref{eq:intsymma}) as
\be
\int{dr'\over r'}K(r,r')y(r')=\int dv\,w_K(v-u){K(e^u,e^{v})y(e^{v})\over
w_K(v-u)} 
\label{eq:oopp}
\ee
where the weight function is chosen to mimic the singular behavior of the
kernel (eq. \ref{eq:sing}) 
\be
w_K(x)\equiv {1\over x^2} + {15\over 8}\log{|x|\over 8} + {179\over 48},
\label{eq:wt}
\ee

Before proceeding further we must soften the kernel to smooth the
singularity. We do this by softening the weight function, that is, we replace
equation (\ref{eq:oopp}) by
\be
\int{dr'\over r'}K(r,r')y(r')=\int dv\,\widetilde w_K(v-u)
{K(e^u,e^{v})y(e^{v})\over w_K(v-u)}, 
\label{eq:oopps}
\ee
Following the prescription in equation (\ref{eq:soften}) the softened weight
function is
\be
\widetilde w_K(x)\equiv {x^2-\beta^2\over (x^2+\beta^2)^2} + {15\over 8}
\log{(x^2+\beta^2)^{1/2}\over 8}+ {179\over 48}.
\label{eq:ooppsoft}
\ee
It should be stressed that this softened kernel does not correspond to any
simple force law between two particles. Its key advantages are that it is
numerically convenient when deriving a quadrature rule (see below); it has the
correct behavior for softened gravity as $x\to0$; and the softened kernel is
still symmetric in $r$ and $r'$ so the operator remains Hermitian. The
softening not only simplifies the numerical evaluation of the integral but
also can be used to mimic the effects of velocity dispersion in collisionless
disks (cf. Fig. \ref{fig:wkb}).

Now if $y(r')$ is smooth, the expression $z(v)\equiv K(e^u,e^v)y(e^v)/
w_K(v-u)$ in equation (\ref{eq:oopp}) should also be smooth, so to evaluate
the contribution to the integral from the interval $[u_k,u_{k+1}]$ we apply a
four-point quadrature rule of the form
\be
\int_{u_{k}}^{u_{k+1}}dvK(e^u,e^v)y(e^v)=\sum_{j=0}^3 W_jz_{k+j-1},
\label{eq:quad}
\ee
where $z_k\equiv z(u_k)$, and $z_{-1}=z_{N+1}=0$. The weights $W_j$ are chosen
so that the integration is exact if $z(v)$ is a cubic polynomial, and can be
evaluated analytically using the integrals $\int_a^b
v^n\widetilde w_K(v-u)dv$, $n=0,\ldots,3$ \nocite{recipes}(Press et al. 1992).

Once the integral in equation (\ref{eq:intsymma}) has been replaced by a sum,
solving the integral equation reduces to finding the eigenvalues and
eigenvectors of a symmetric $N\times N$ matrix, which can be done by standard
methods (Press et al. 1992). 

A similar strategy is used to evaluate the precession rate: we write the
integral in equation (\ref{eq:precess}) as
\be
\int{dr'\over r'}Q(r,r')=\int dvw_Q(v-u){Q(e^u,e^{v})\over w_Q(v-u)}
\label{eq:wwrr} 
\ee
where
\be
w_Q(x)\equiv {1\over x^2} + {3\over 8}\log{|x|\over 8} + {11\over 48} + \half
x^2; 
\label{eq:wtq}
\ee
the term  $\half x^2$ is added to eliminate a zero of $w_Q(x)$ that
would otherwise occur at $|x|=1.6712$, but has no effect near the singularity
at $x=0$. We then soften $w_Q(x)$ to $\widetilde w_Q(x)$ and recompute the
weights $W_j$ for the quadrature rule. 

\section{Slow modes of disk models}

\noindent
In our numerical calculations we typically use grids with $N=400$--800. All
the significant digits that we quote should be correct in the limit
$N\to\infty$.

Numerical mode calculations are more difficult with unsoftened gravity than
with softened gravity: the diagonal matrix elements become large when
the softening is much less than the grid size, so the accuracy is degraded by
roundoff error. Typically the minimum practical softening parameter was
$\beta=10^{-4}$. 

Our numerical method always yields $N$ eigenvectors and eigenvalues, but many
of these are singular (i.e. the amplitude is much larger at a single grid
point than elsewhere). These correspond to the singular (van Kampen) modes
that occupy the continuous region of the eigenvalue spectrum.

\subsection{The Kuzmin disk}

\label{sec:kuz}

\noindent
We use units in which $G=M=a=1$ (cf. \S \ref{sec:kuzmin}). We usually employ a
grid that covers the range $-7\le\log r\le 5$. To account for precession due
to the mass inside the innermost grid point, we divide the disk into ``live''
and ``frozen'' components: the frozen component is a Kuzmin disk with $\log
a=-5$ and the same central surface density as the original disk, while the
live component is the difference between the original and frozen disks. This
approximation is legitimate if the amplitude of the normal mode is negligible
for $\log r\lesssim -5$. We have checked that our results are not sensitive to
the size of the frozen disk.

Since softened self-gravity is the only collective effect, the shape of a mode
is independent of the disk mass and the eigenfrequency is proportional to disk
mass, so we are free to set $M_d=1$ as well, even though the calculations are
only valid in real disks if $M_d\ll M$.

We classify the modes as g-modes and p-modes, as described in \S\ref{sec:wkb}. 

\begin{figure}
\plotone{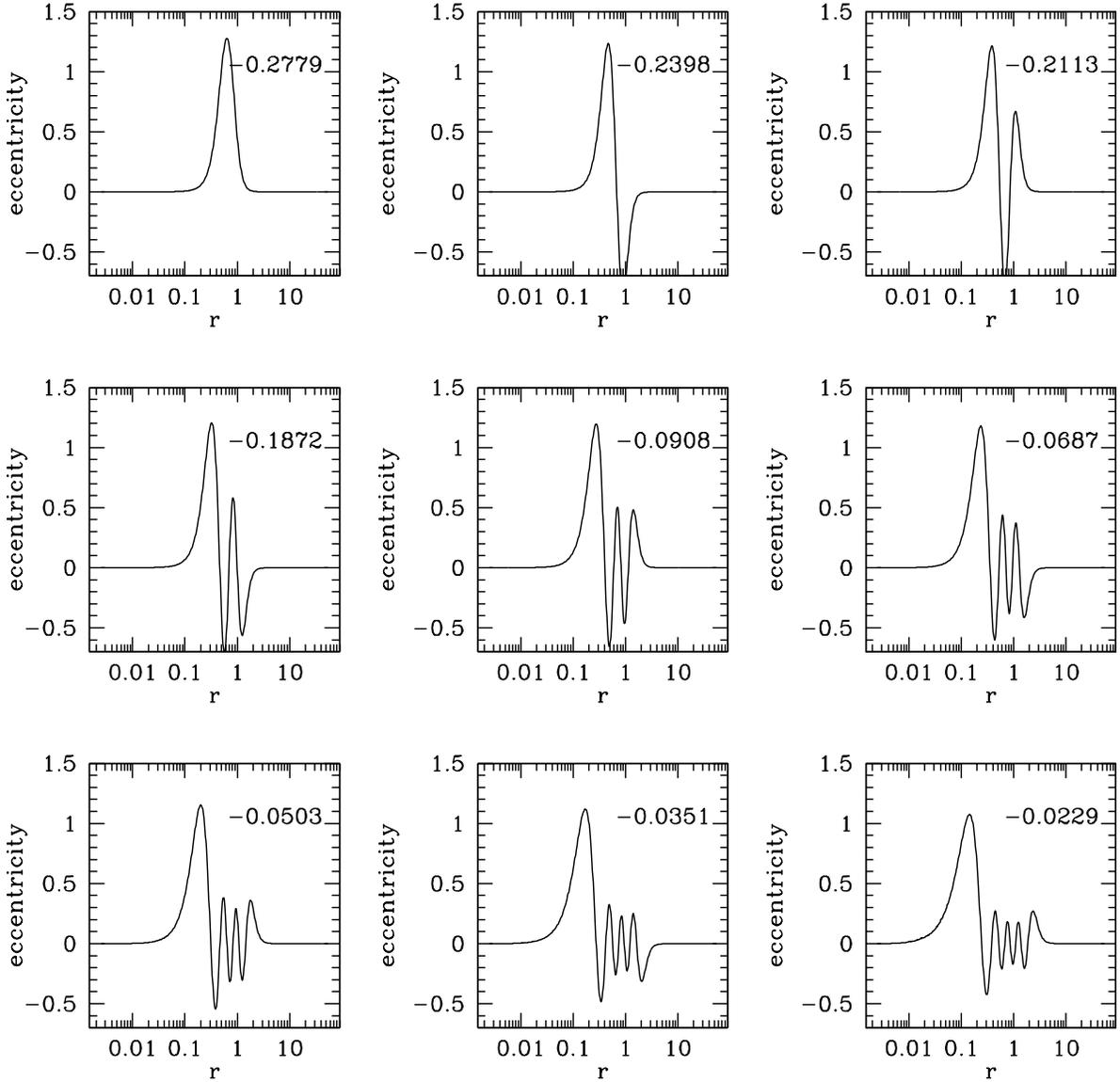}
\caption{Slow g-modes in the Kuzmin disk with $\lambda=0.1$ and
$\beta=0$. The modes are sorted by the number of nodes. The normalization is
chosen so that $\int e_K^2(r)d\log r=1$, where the eccentricity $e_K(r)$ is
defined in equation (\ref{eq:eccdefa}). The number in each panel is the
eigenfrequency, in units where $G=M=M_d=a=1$. The eigenfrequency scales as
$(M_d/M)(GM/a^3)^{1/2}$.}
\label{fig:shape}
\end{figure}

\subsubsection{g-modes}

\noindent
We first examine Kuzmin disks with $\lambda=0.1$ and $\beta\simeq0$ (actually
$\beta=10^{-4}$), corresponding to the case in which 90\% of the precession is
due to an external source. The reasons for examining this somewhat artificial
case are that (i) these disks do not support p-modes, since gravity is
unsoftened; (ii) g-modes are accurately described by the WKB approximation
since $\lambda\ll1$. Thus we can both isolate the g-modes and compare their
behavior to the analytic discussion in \S \ref{sec:wkb}.

This disk exhibits a rich spectrum of discrete slow modes, which can be sorted
by the number of nodes; the modes with $\le 8$ nodes are shown in Figure
\ref{fig:shape} along with their eigenfrequencies. As expected from the WKB
analysis, the modes in Figure \ref{fig:shape} all have $\omega<0$, and the
modes with the fewest nodes have the most negative values of $\omega$.

\begin{figure}
\plottwo{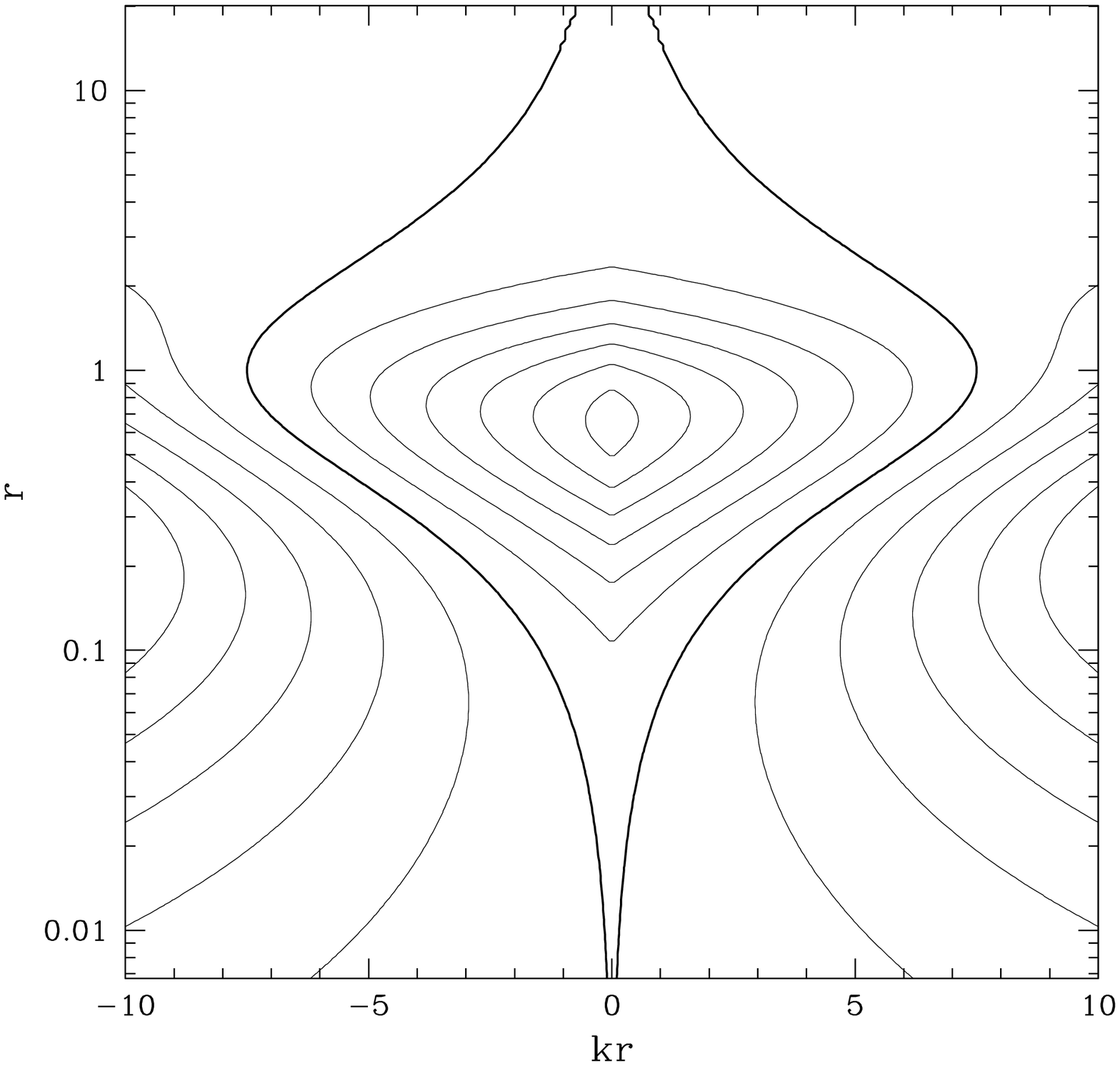}{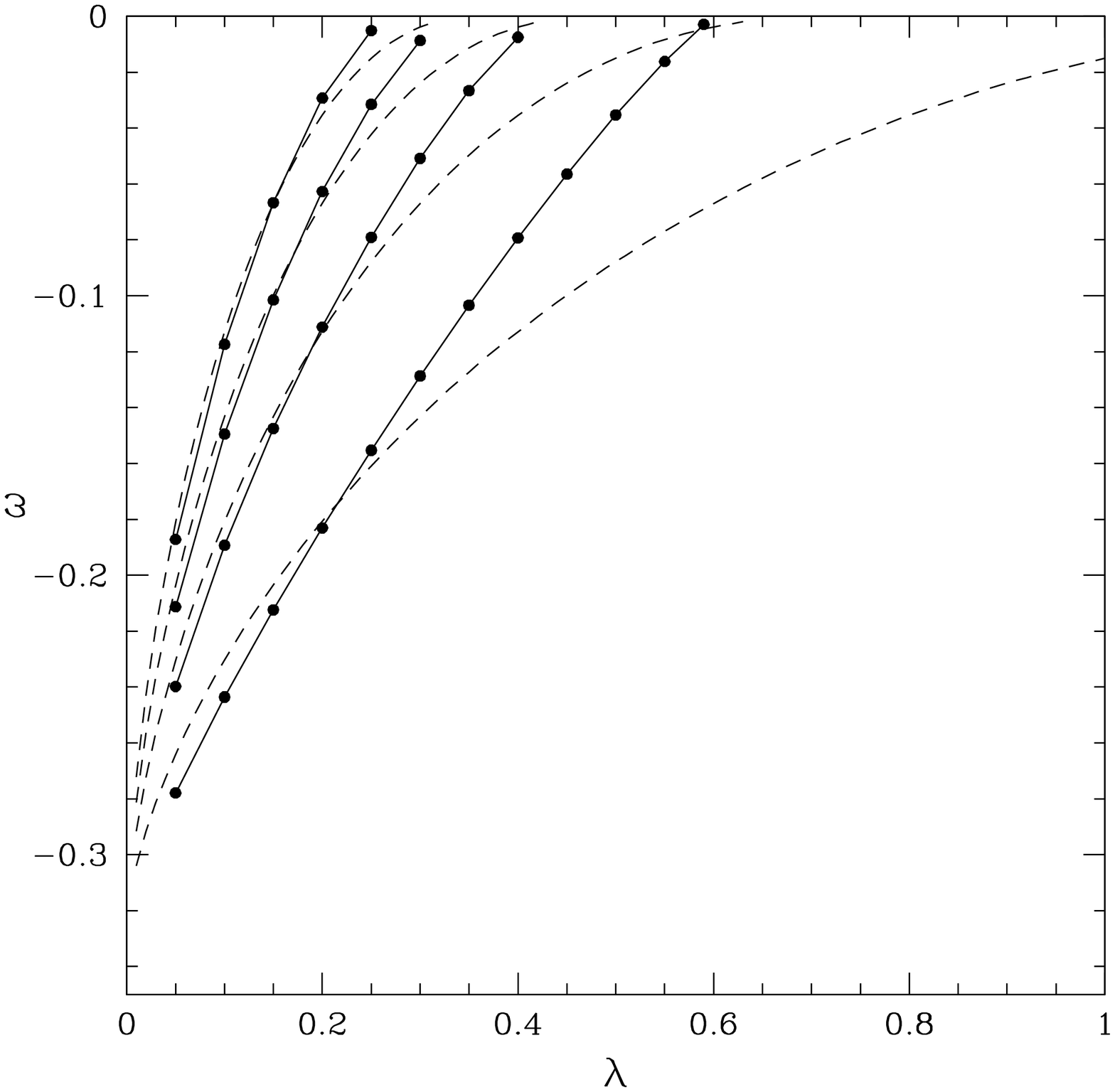}
\caption{(Left) Contours of constant frequency $\omega$ for WKB slow waves in a
Kuzmin disk with $\lambda=0.1$ and $\beta=0$, as a function of wavenumber $kr$
and radius $r$. The contours are plotted at intervals of 0.05; the heavy
contour is $\omega=0$. The g-modes correspond to the closed contours. (Right)
The eigenfrequencies of g-modes with 0, 1, 2, or 3 nodes (bottom to top) for
the Kuzmin disk, as a function of $\lambda$ (solid circles and lines). The
dashed lines show the predictions of the WKB approximation
(eqs. \ref{eq:wkbff} and \ref{eq:wkbgg}). There are no slow g-modes for
$\lambda\ge0.6$.}
\label{fig:wkbk}
\end{figure}

The eigenfrequencies can be predicted using the WKB approximation. The 
dispersion relation (\ref{eq:wkb}) can be modified to include a factor
$\lambda$ that accounts for precession from an external source
(cf. eq. \ref{eq:intsymma}), and rewritten as
\be
|k(r)|={\Omega(r)\over\pi G\lambda\Sigma_d(r)}[\omega-\dot\varpi(r)].
\label{eq:wkbff}
\ee
The frequency contour diagram analogous to Figure \ref{fig:wkb} is shown in
the left panel of Figure \ref{fig:wkbk}. The g-modes correspond to the 
contours centered on $kr=0$, $r=r_0=0.6547$; they have turning points at radii
$r_\pm$ defined by $\dot\varpi(r_\pm)=\omega$. The eigenfrequencies satisfy
the quantum condition that the total phase change around the contour is an
integer multiple of $2\pi$,
\be
2\int_{r_-}^{r_+}|k(r)|dr=2\pi n, \qquad n=1,2,3,\ldots
\label{eq:wkbgg}
\ee
(there are additional phase shifts of $\pm\half\pi$ at the turning points, but
these contributions cancel; see, for example, \nocite{shu90}Shu et
al. 1990). The right panel of Figure
\ref{fig:wkbk} shows the eigenfrequencies of the modes with 0, 1, 2, or 3
nodes, as a function of $\lambda$, along with the WKB approximation from
equations (\ref{eq:wkbff}) and (\ref{eq:wkbgg}) for $n=1,\ldots,4$. We see
that the WKB approximation predicts the eigenfrequency fairly well for
$\lambda\lesssim 0.3$. At larger $\lambda$, the WKB predictions are seriously
in error.

As $\lambda$ increases, the frequency of each g-mode declines in absolute
value, until the mode terminates when its frequency reaches zero.  There are
{\em no} g-modes in a Kuzmin disk with unsoftened gravity for $\lambda\ge
0.6$. In particular, an isolated Kuzmin disk ($\lambda=1$) does not support
g-modes. The reason for this result can be understood qualitatively.  As
$\lambda$ increases (i.e. the importance of external precession declines),
$k(r)$ declines as $\lambda^{-1}$, (eq. \ref{eq:wkbff}) so that the integral
on the left side of equation (\ref{eq:wkbgg}) also declines. Eventually this
integral is less than $2\pi$ even for the largest closed contour in Figure
\ref{fig:wkb}, at which point the disk cannot support any g-modes.

\begin{figure}
\plotone{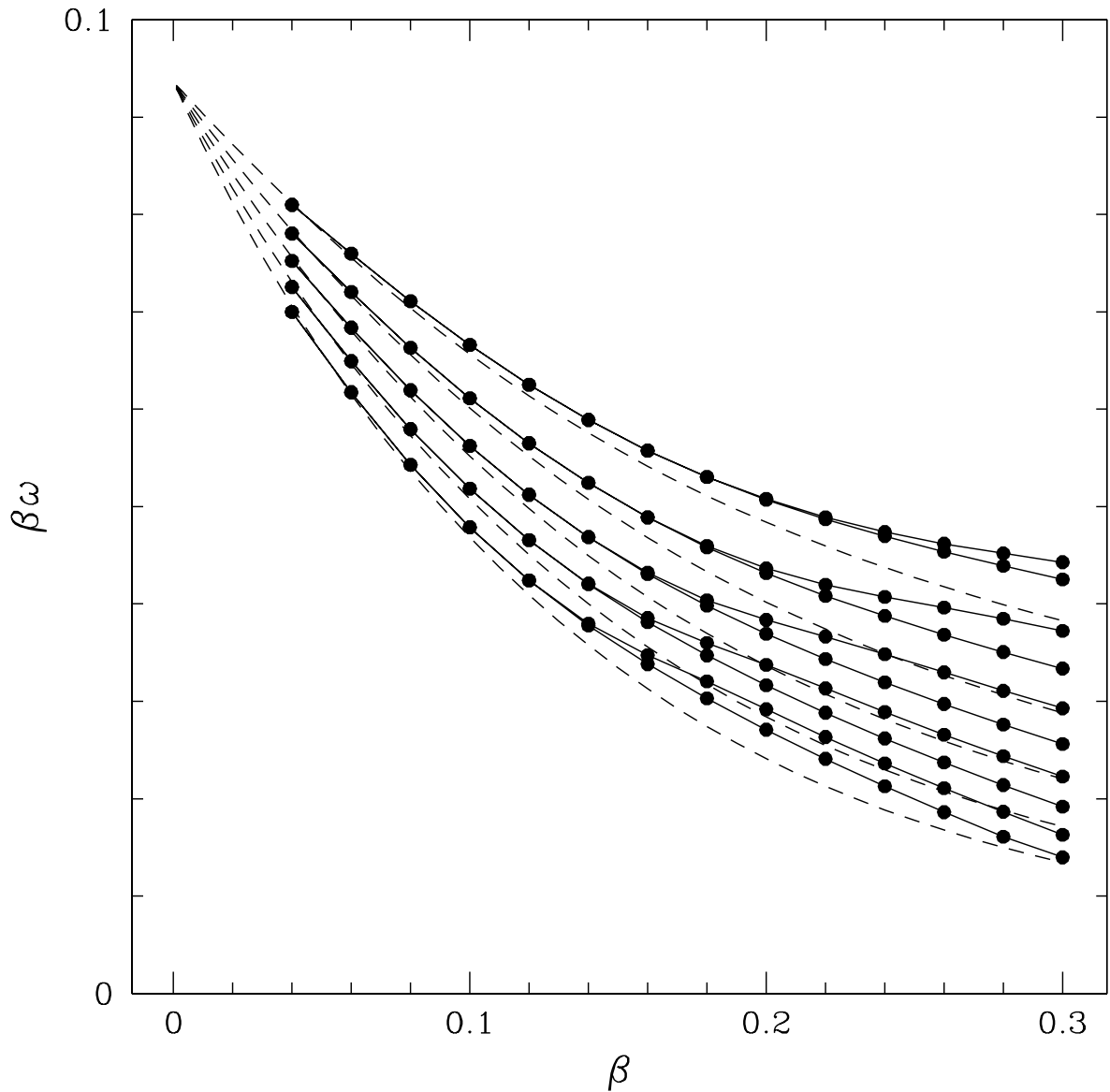}
\caption{Eigenfrequencies of slow p-modes in Kuzmin disks with softened
gravity. The units are $G=M=M_d=a=1$. The filled circles and solid curves show
numerical solutions for the ten highest eigenfrequencies. The dashed lines
show the predictions from the WKB approximation (\ref{eq:wkbpp}) for
$n=1,\ldots,5$. In this approximation the modes appear in degenerate
leading/trailing pairs. We have plotted $\beta\omega$ instead of $\omega$ to
expand the vertical scale (cf. eq. \ref{eq:limit}).}
\label{fig:ppp}
\end{figure} 

\subsubsection{p-modes}

\noindent
Next we examine p-modes in isolated Kuzmin disks with softened gravity. Here
the WKB analysis is useful even for isolated ($\lambda=1$) disks so long as
$\beta\ll1$, since the characteristic wavenumber is $k_0=1/b=1/\beta r$. The
WKB analysis predicts that p-modes have $\omega>0$; that the modes with the
fewest nodes have the largest $\omega$; that the modes occur in degenerate
pairs; and that the quantum condition is
\be
2\int_{r_-}^{r_+}[k_+(r)-k_-(r)]dr=2\pi (n-\half), \qquad n=1,2,3,\ldots
\label{eq:wkbpp}
\ee
Here $k_-<k_0<k_+$ are the two solutions of the dispersion relation
(\ref{eq:dispsoft}); $r_-$ and $r_+$ are the turning points at which
$k_-=k_+$ for a given $\omega$-contour; and the term $\half$ on the right side 
arises because there is a phase shift of $\half\pi$ at each turning point
\nocite{shu90}(Shu et al. 1990). 

\begin{figure}
\plotone{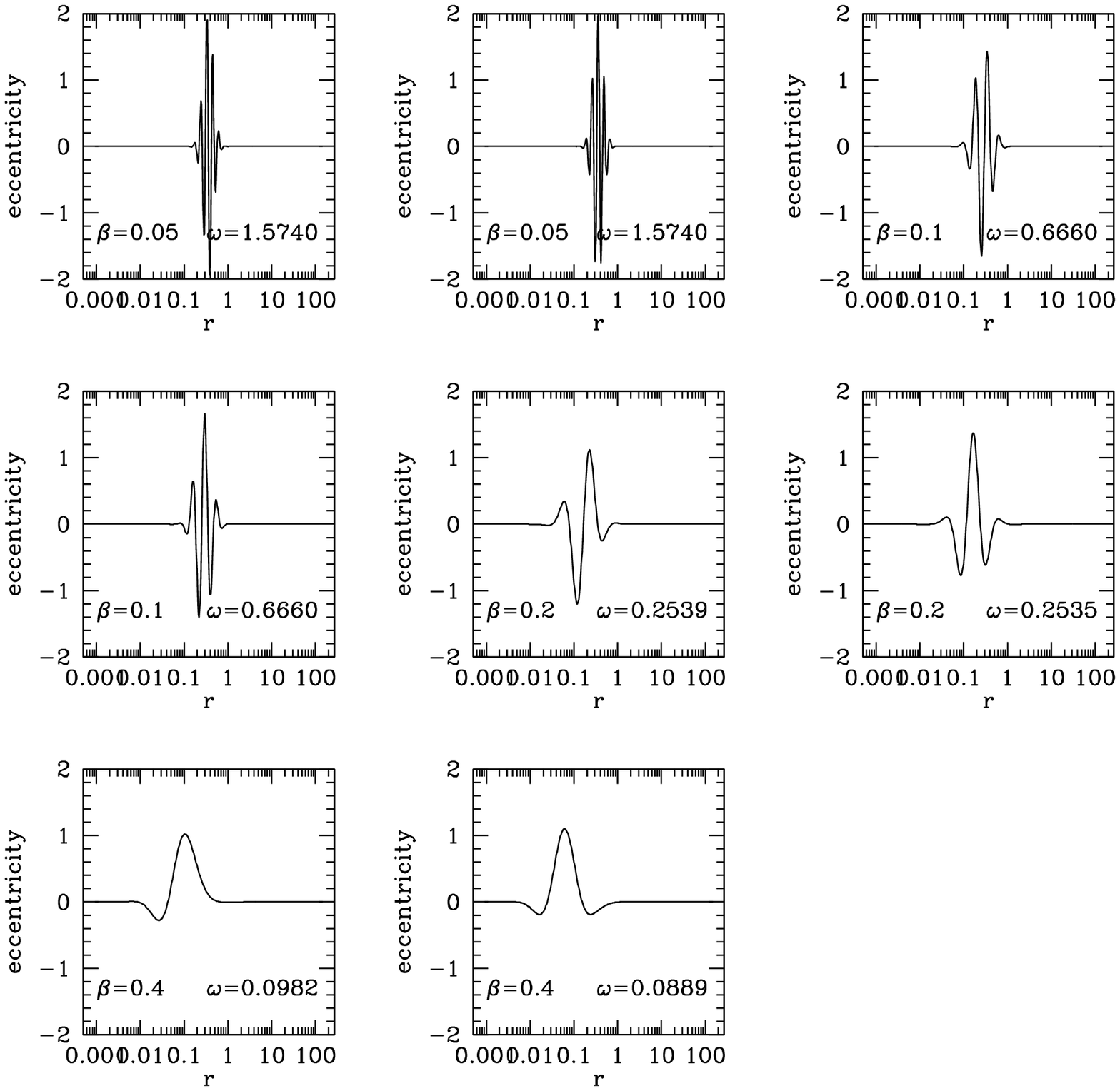}
\caption{Slow p-modes in the isolated Kuzmin disk with softened gravity.
The two modes with the highest frequency are shown for disks with
$\beta=0.05$, $0.1$, $0.2$, and $0.4$. Note that for $\beta\ll1$ the
frequencies are nearly degenerate. The normalization is chosen so that $\int
e_K^2(r)d\log r=1$, where the eccentricity $e_K(r)$ is defined in equation
(\ref{eq:eccdefa}).}
\label{fig:shapp}
\end{figure}

Figure \ref{fig:ppp} shows the predictions of this quantum condition for the
Kuzmin disk, as a function of the softening parameter $\beta=b/r$. As
$\beta\to0$ the eigenfrequencies diverge as $\beta^{-1}$, so we have plotted
$\beta\omega$. The limit point is
\be
\lim_{\beta\to 0}\beta\omega={5^{5/4}\over
2^{5/2}3^{3/2}e}=0.093575. \label{eq:limit} 
\ee
We have also plotted the ten largest eigenfrequencies from numerical
calculations of the modes. These agree very well with the WKB approximation
for small $\beta$, and even for $\beta=0.3$ the WKB estimates of the
eigenfrequencies are accurate within $\sim 20\%$. Figure \ref{fig:shapp} shows
the shape of the fundamental ($n=1$) modes as a function of $\beta$. In
contrast to the g-modes, the fundamental p-modes can have many nodes, because
they are modulated at the high spatial frequency $k_0$.

\begin{figure}
\plotone{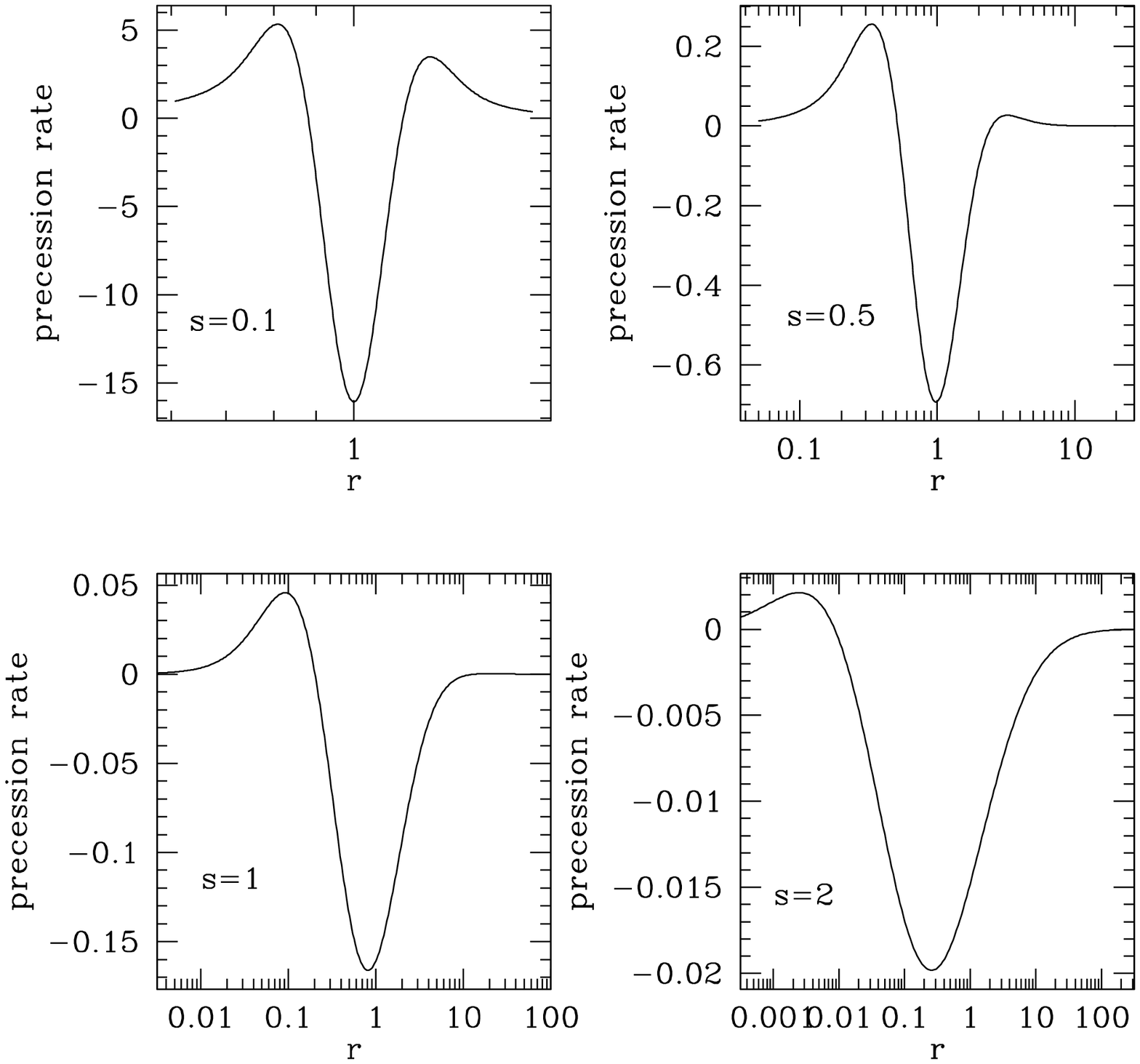}
\caption{The apsidal precession rate $\dot\varpi$ for Gaussian rings with
surface-density distribution described by equation (\ref{eq:gauss}).}
\label{fig:preg}
\end{figure}

\subsection{Gaussian rings}

\noindent
In this Section we examine slow modes in rings, by which we mean disks whose
surface density approaches zero as $r\to\infty$ and as $r\to 0$. As a model
for rings we take the surface-density distribution
\be
\Sigma_d(r)={K\over r}\exp\left[-(\log r)^2\over 2s^2\right];
\label{eq:gauss}
\ee
where the constant $K$ is related to the disk mass by
$M_d=(2\pi)^{3/2}Ks\exp(-\half s^2)$. The precession rate in these rings is
shown in Figure \ref{fig:preg}. Like the Kuzmin disks, Gaussian rings exhibit
a single minimum in the precession rate, but the central minimum is surrounded
by shoulders in which the precession rate is positive (prograde).

\begin{figure}
\plotone{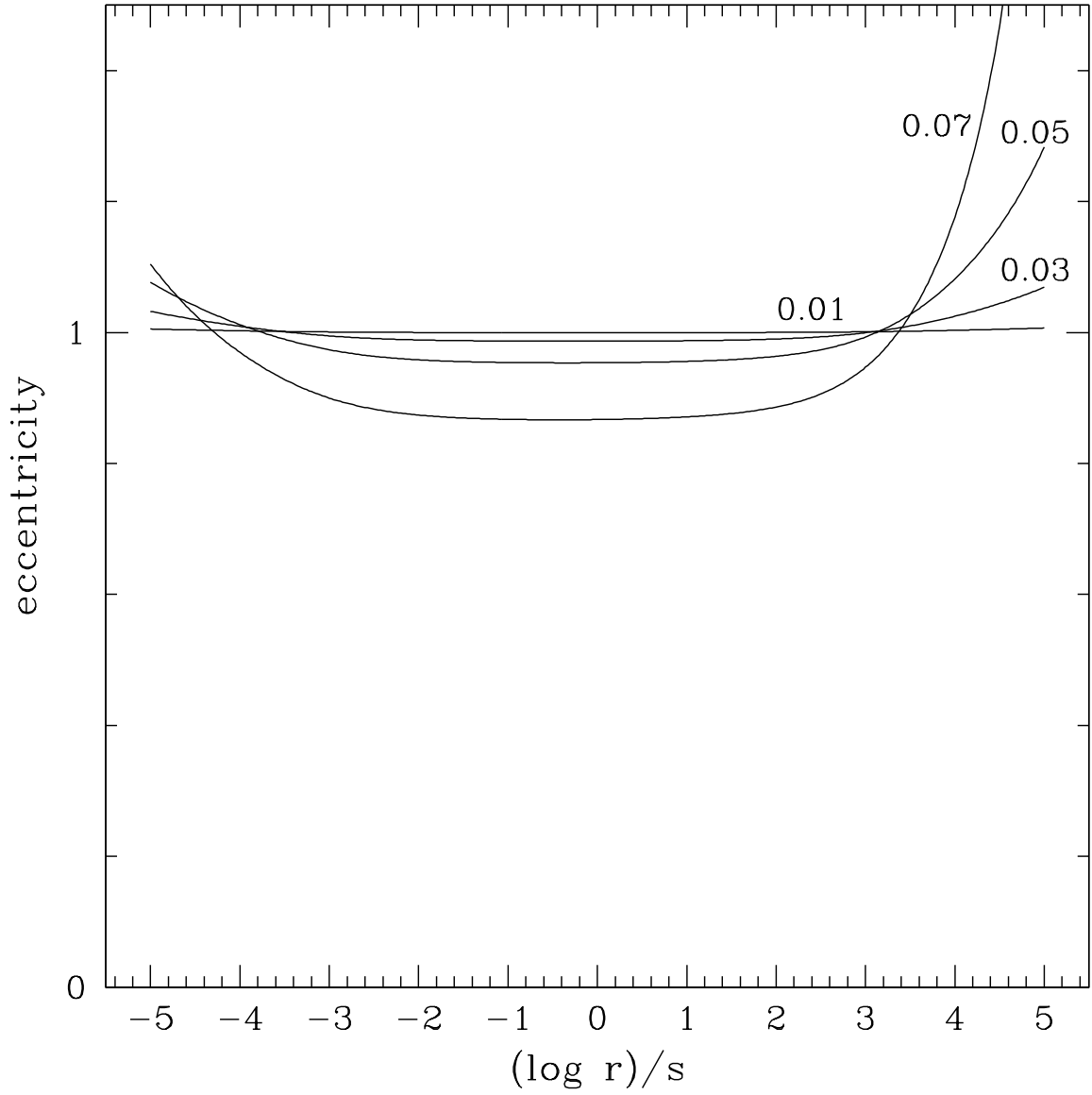}
\caption{The fundamental slow modes for narrow Gaussian rings
(eq. \ref{eq:gauss}). The modes are normalized so that $\int e^2(r)d\log
r/\int d\log r=1$, and labeled by $s$, the dispersion in $\log r$. There are
no slow modes for $s> 0.07$.}
\label{fig:narrow}
\end{figure}

\subsubsection{Narrow rings}

\noindent
We showed at the end of \S \ref{sec:eigen} that there is a trivial slow mode
for narrow rings, with zero eigenfrequency and constant velocity
perturbation. We can extend this analytic result with numerical solutions of
the eigenvalue equation for narrow Gaussian rings. 

We find that slow modes for isolated rings with unsoftened gravity are present
only if the ring is rather narrow, $s\lesssim 0.07$. Only the fundamental mode
is non-singular; these are shown in Figure \ref{fig:narrow}. For $s\lesssim
0.03$ the shape of the fundamental mode is nearly flat, as expected from the
trivial solution. At larger $s$ the amplitude in the outer parts of the ring 
grows more and more sharply, until for $s\gtrsim 0.07$ we were unable to find
any non-singular modes. The frequency of the fundamental mode declined from
$\omega=1.1074$ at $s=0.01$ to $\omega=0.6405$ at $s=0.07$, in units where the
ring mass $M_d=1$.

\begin{figure}
\plotone{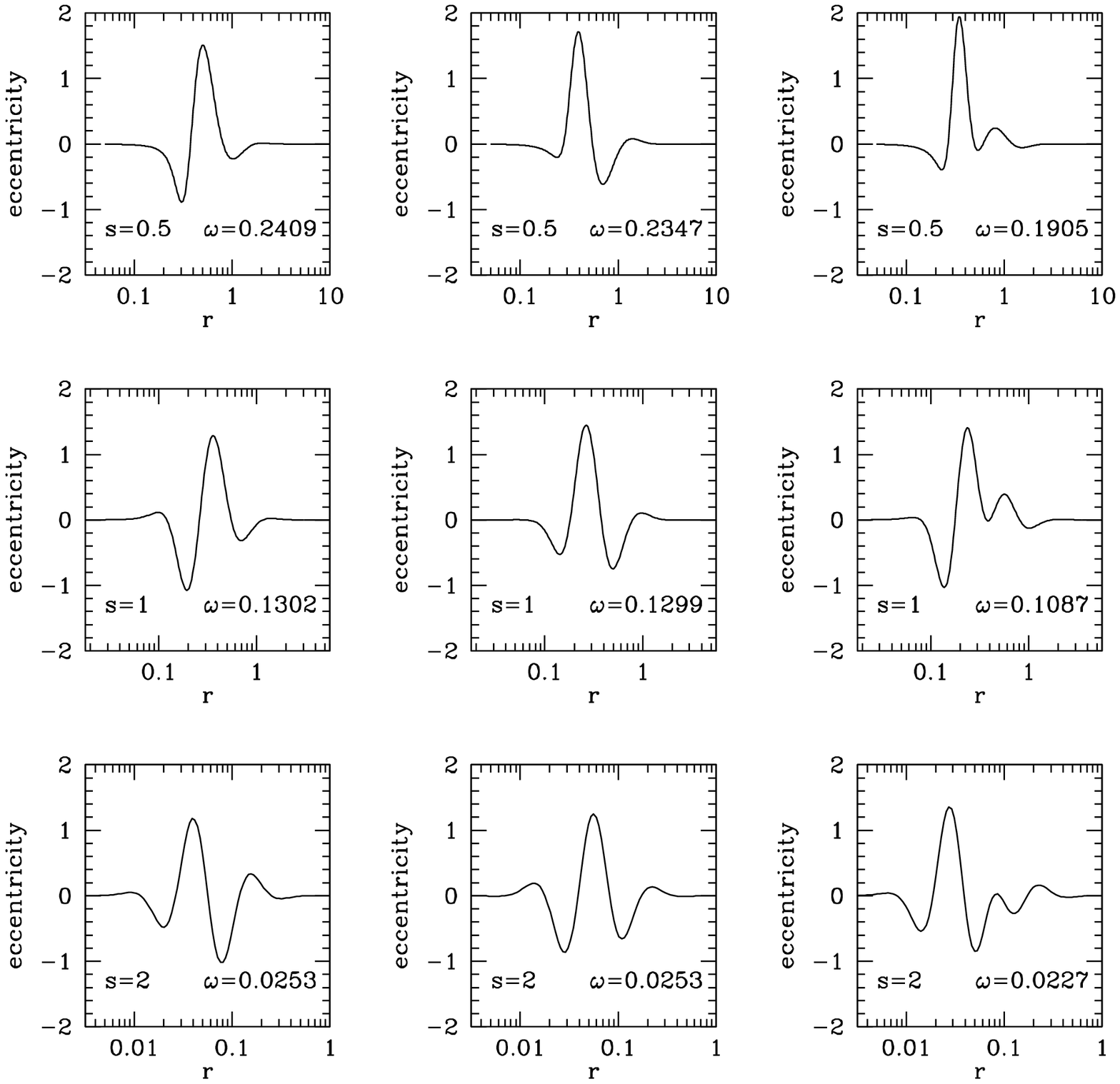}
\caption{Slow p-modes in the isolated Gaussian ring (eq. \ref{eq:gauss}) with
softened gravity, $\beta=0.2$. The panels are labeled with the width parameter
$s$ and the eigenfrequency. The scale in $\log r$ is the same in all panels
but the origin is different.}
\label{fig:gauss}
\end{figure}

\subsubsection{Broad rings}

\noindent
We have examined a sequence of broader Gaussian rings, with softening
parameter $\beta=0.2$ and width $s=0.5,1,2$. The three highest-frequency modes
for each ring are shown in Figure \ref{fig:gauss}. These are all p-modes.

\begin{figure}
\plotone{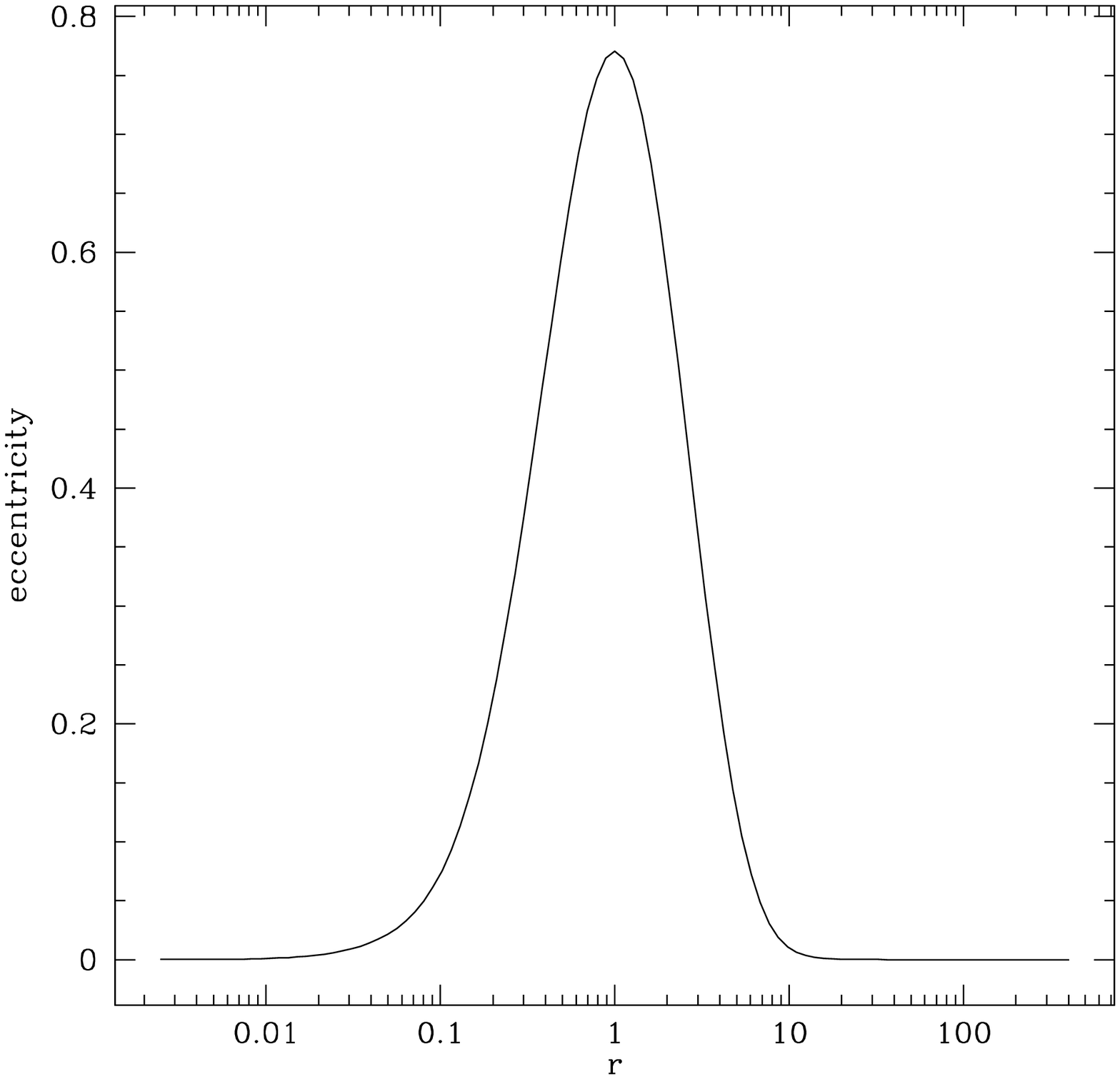}
\caption{The fundamental g-mode in a Gaussian ring with width $s=1$ and
$\lambda=0.5$. Gravity is unsoftened ($\beta=10^{-4}$). The frequency of the
mode is $\omega=-0.1527$ and is proportional to the disk mass, which is here
chosen to be $M_d=1$.}
\label{fig:gaussg}
\end{figure}

Like the Kuzmin disk, isolated broad Gaussian rings with unsoftened gravity do
not support g-modes. We have examined a sequence of Gaussian rings with $s=1$,
near-zero softening ($\beta=10^{-4}$), and increasing $\lambda$. The
fundamental g-mode is well-defined for $\lambda=0.5$ (Figure
\ref{fig:gaussg}), but disappears by $\lambda=0.59$.

\begin{figure}
\plottwo{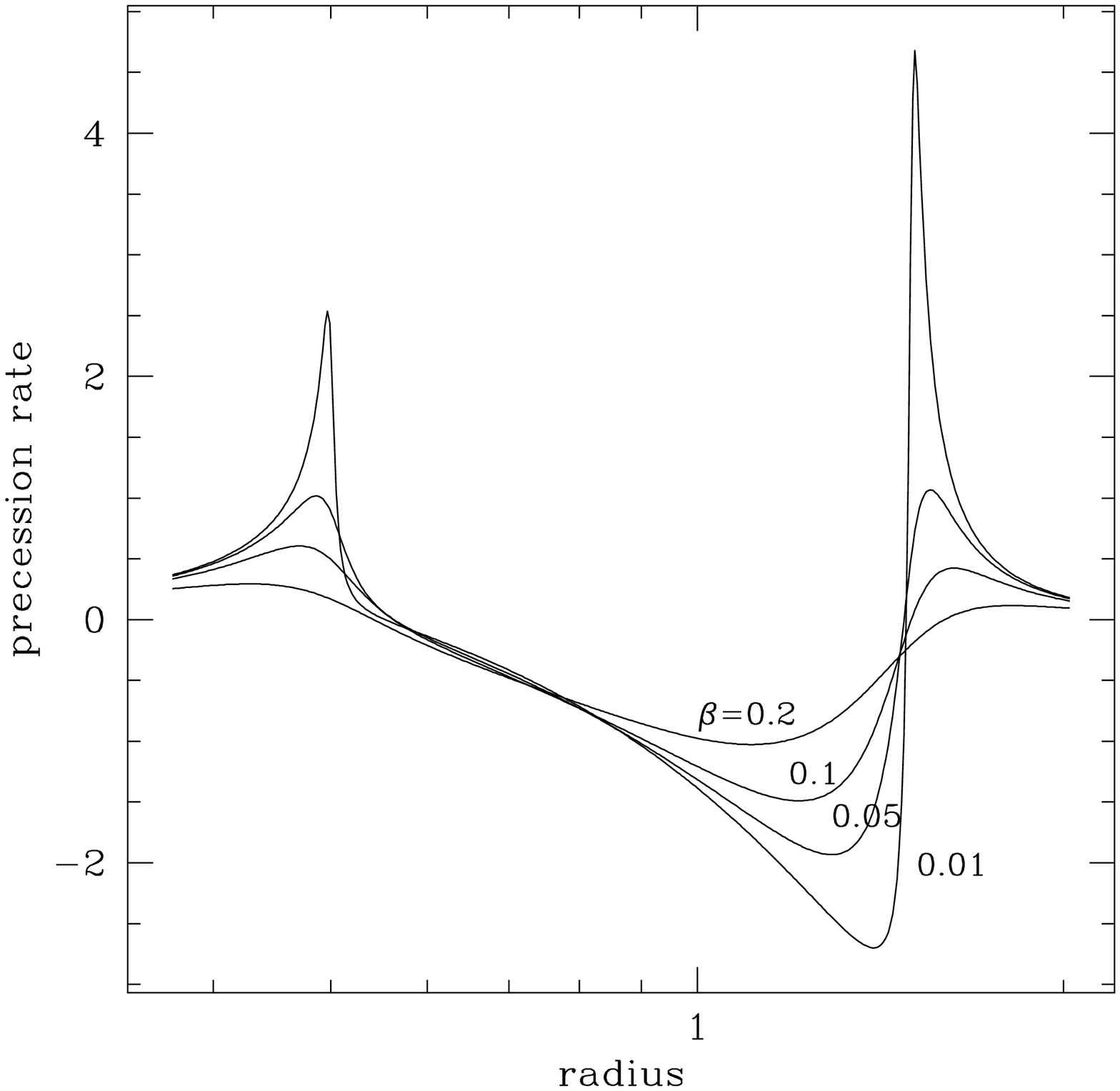}{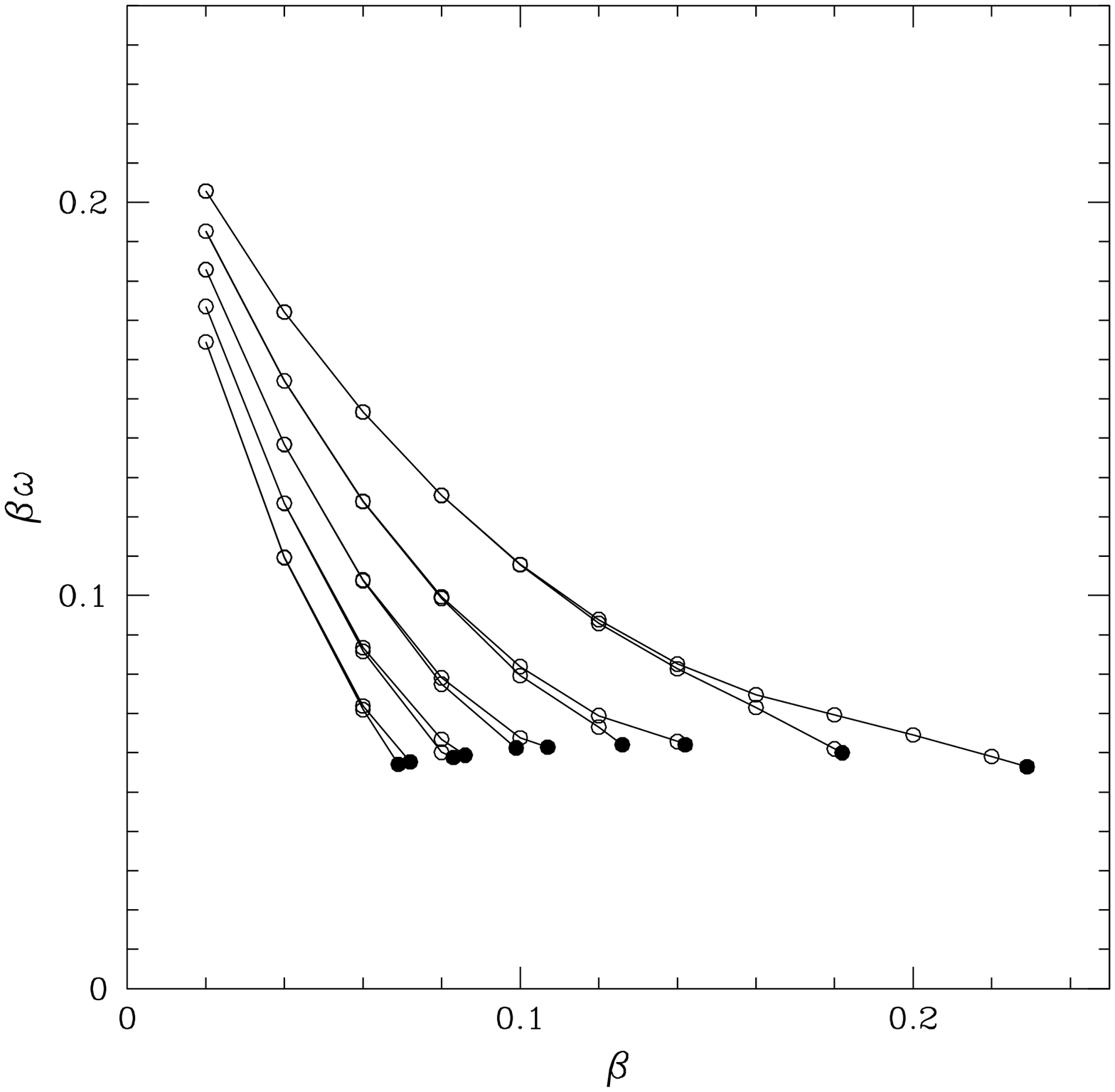}
\caption{(Left) The apsidal precession rate for the Jacobs-Sellwood ring
(eq. \ref{eq:sell}) for softening parameter $\beta=0.01,0.05,0.1,0.2$. (Right)
Slow p-modes in the isolated Jacobs-Sellwood ring with softened gravity. We
show the ten modes with the highest frequency. The modes occur in degenerate
pairs when $\beta\ll1$. Each mode terminates at a maximum softening parameter,
shown by a filled circle. The units are chosen so that $G=M=M_d=1$, and
$\omega\propto M_d$.}
\label{fig:sell}
\end{figure}

\subsection{The Jacobs-Sellwood ring}

\noindent
\nocite{js99} Jacobs and Sellwood (1999) have reported two-dimensional $N$-body
simulations of a low-mass annular disk orbiting a central mass. They were able
to establish long-lived $m=1$ normal modes that survived with little sign of
decay for over 1500 orbital periods. The surface-density distribution in the
disk was
\be
\Sigma_d(r)={2M_d\over \pi}\left[1-4(1-r)^2\right]^{1/2}, \qquad
\half<r<\case{3}{2},
\label{eq:sell}
\ee
and zero otherwise. In contrast to our other models, this disk has sharp edges
at $r_\pm={3\over2},{1\over2}$, where the surface density goes to zero as
$\Sigma_d\propto |r-r_\pm|^{1/2}$. To ensure that our numerical methods handle
edge effects properly, we extend the ring to $fr_+$ and $r_-/f$, $f=1.35$,
with a very low surface-density envelope, $\Sigma_d=10^{-8}M_d$.

\begin{figure}
\plotone{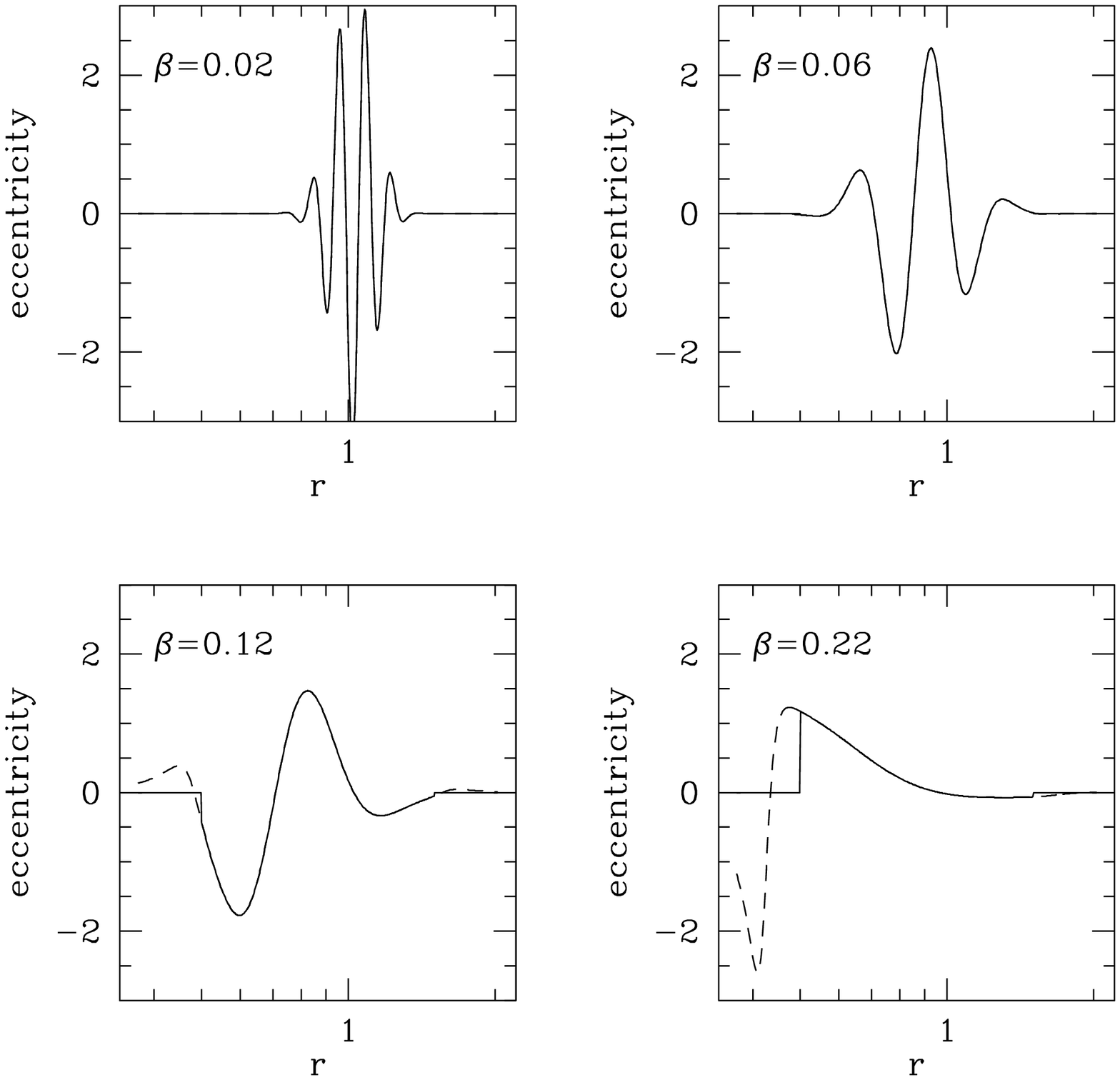}
\caption{The p-mode in the Jacobs-Sellwood ring with the highest frequency,
for $\beta=0.02,0.06,0.12,0.22$. The dashed lines show the amplitude of the
mode in the a low surface-density envelope ($\Sigma_d=10^{-8}M_d$) that
surrounds the ring in our numerical model.}
\label{fig:sellmode}
\end{figure}

The precession rate in the unperturbed disk (\ref{eq:sell}) is shown in the
left panel of Figure \ref{fig:sell}, for several values of the softening
parameter $\beta$. The cusps at the inner and outer disk edge $r_\pm$ arise
because of the sharp edges.

The right panel of Figure \ref{fig:sell} shows the frequencies of p-modes as a 
function of the softening parameter. The ten modes with the largest frequency
are shown. The modes occur in degenerate pairs for $\beta\ll1$ and the
degeneracy is gradually lifted as $\beta$ increases. Each mode terminates at a 
maximum softening parameter, which is shown by a filled circle. For
$\beta>0.229$ there are {\it no} non-singular modes. 

Figure \ref{fig:sellmode} shows the shape of the highest frequency
(fundamental) p-mode for several values of the softening parameter. Note that
for the larger softening parameters, the amplitude of the mode does not go to
zero at the disk edge. 

\section{Earlier work}

\label{sec:other}

\noindent
\nocite{srid99}Sridhar, Syer and Touma (1999) recognized the importance of slow
modes in nearly Keplerian disks, and both formulated and solved numerically
the linear eigenvalue problem. However, they employed the equations of
Laplace-Lagrange secular perturbation theory \nocite{mur99}(Murray and Dermott
1999), which is designed for planetary systems and so approximates the disk as
a set of non-intersecting rings. This approach does not give the correct
answers for a continuous disk, even as the number of rings becomes arbitrarily
large, because it assumes that the eccentricities are small compared to the
distance between rings. For example, all of the eigenfrequencies are
non-negative in Laplace-Lagrange theory (Sridhar et al. 1999), whereas we have
found negative eigenfrequencies in both the WKB approximation and numerical
solutions.

\nocite{lg99} Lee and Goodman (1999) have investigated stationary linear and
nonlinear $m=1$ density waves in nearly Keplerian disks. They have derived the
nonlinear dispersion relation in the tight-winding approximation, and found
self-similar logarithmic spirals with zero frequency in the special case where
the disk surface density $\Sigma_d(r)\propto r^{-3/2}$. In the linear limit,
their self-similar spirals exhibit properties reminiscent of the results of
this paper: isolated disks do not support long waves, and zero-pressure disks
in an external potential only support waves if the parameter $\lambda$ from
equation (\ref{eq:intsymma}) satisfies $\lambda<[\Gamma({1\over
4})/\Gamma({3\over 4})]^4/144=0.53214$, similar to our finding that there are
no g-modes in the Kuzmin disk or $s=1$ Gaussian ring if $\lambda\gtrsim
0.5$--0.6.

\nocite{st99}Sridhar and Touma (1999) have examined the slow dynamics of
eccentric orbits in nearly Keplerian potentials with a non-axisymmetric
potential perturbation. They found two general families of orbits, lenses and
loops, and suggest that a self-consistent lopsided disk around a point mass
could be assembled from aligned loop orbits. Our results extend this
conclusion by showing that aligned loops generally can support self-consistent
slow g-modes, but only if the precession is dominated by an external
non-Keplerian potential or the disk is a narrow ring. In essence, the aligned
loop orbits discussed by Sridhar and Touma work {\em too} well: the
non-axisymmetric gravitational field that they produce is too strong for a
self-consistent slow mode. The progeny of aligned loop orbits, which
\nocite{st99}Sridhar and Touma (1999) call librating loop orbits, can support
self-consistent p-modes in isolated disks, but large-scale p-modes require
that the velocity dispersion is large.

\nocite{sta99}Statler (1999) has constructed sequences of periodic orbits in
the potential of a point mass plus a lopsided disk. He argues that the
precession of a self-consistent lopsided disk must be prograde ($\omega>0$),
and that the eccentricity $e_K(r)$ must have a node. We find that neither of
these results is generally true for g-modes: for example, the fundamental
g-mode in the Kuzmin disk with $\lambda=0.1$ (Fig. \ref{fig:shape}) has no
nodes and $\omega<0$. Nevertheless, Statler's arguments may well describe the
essential dynamics of some lopsided disks. 

\nocite{gt79} Goldreich and Tremaine (1979) investigated slow modes in
planetary rings. The thrust of their investigation was quite different, for
several reasons: there was strong differential precession due to an external
source (the planet's quadrupole moment), so that uniform precession was
sustained only by a strong eccentricity gradient ($de/d\log a\gg1$); the ring
was very narrow ($\Delta a/a\sim 10^{-4}$); the ring was sharp-edged, so that
the ``resonant cavity'' involved reflection at the disk edges rather than at
turning points in the dispersion relation; and finally, later work
\nocite{cg00} (Chiang and Goldreich 2000) shows that the mode properties in
narrow, dense, particulate rings are strongly affected by collisions near the
disk edge. In contrast, the slow modes examined here should be insensitive to
the details of the disk edges since they are restricted to the interior of the 
disk (except for some of the modes in the strongly softened Jacobs-Sellwood
ring, Figure \ref{fig:sellmode}). 

\section{Conclusions}

\noindent
Disks orbiting massive bodies support slow modes; these are linear $m=1$
normal modes whose eigenfrequency is proportional to the strength of
collective effects in the disk, rather than to its characteristic radial or
azimuthal frequencies. The most important collective effects are likely to be
the self-gravity of the disk, and the velocity dispersion in collisionless
disks or the pressure in fluid disks. The first two of these---and to a lesser
extent the third---can be approximated by softened self-gravity
(Fig. \ref{fig:wkb}). Slow modes are important because these are likely to be
the only large-scale or long-wavelength modes, and hence can dominate the
overall appearance of the disk. Moreover, slow modes are relatively immune to
damping by viscosity, Landau damping, collisions, or other dissipative
mechanisms.

Slow modes in which the disk self-gravity is the dominant collective effect
are described by the same equations used in the classic Laplace-Lagrange
secular perturbation theory for planetary systems, except that (i) the number
of planets $N$ goes to infinity and the mass per planet $m\propto N^{-1}$;
(ii) the Laplace-Lagrange equations are valid in the limit where the planets'
radial excursions are much less than their radial separation, whereas in
continuous disks the radial separation of the mass elements is smaller than
the radial excursions. The simplest way to modify Laplace-Lagrange theory to
account for this difference is to soften gravity.

We have derived the eigenvalue equation (\ref{eq:intsymma}) that describes
slow modes in a nearly Keplerian disk with softened gravity. We have solved
this equation numerically to investigate the properties of modes in the Kuzmin 
disk (eq. \ref{eq:kuzmin}), the Gaussian ring (eq. \ref{eq:gauss}), the
Jacobs-Sellwood ring (eq. \ref{eq:sell}), and a variety of other disks not
reported here. Many of our results can be interpreted in the framework of the
WKB approximation (\S \ref{sec:wkb}). Our conclusions can be summarized as
follows:

\begin{enumerate}

\item All slow modes have real frequency, and thus are stable.

\item There are two main types of discrete slow mode: (i) g-modes involve
long leading and long trailing waves. Their properties are insensitive to the
softening parameter $\beta$ once it is sufficiently small. In general these
are retrograde (eigenfrequency $\omega<0$). The fundamental g-mode has zero
nodes, the lowest eigenfrequency, and a scale comparable to the disk
radius. (ii) The properties of p-modes depend on the softening parameter; in
particular their characteristic wavenumber is $k\sim\pm\beta/r$. In the WKB
approximation that is valid for $\beta\ll1$, p-modes come in degenerate pairs,
one composed of short and long trailing waves and the other of short and long
leading waves. In general p-modes are prograde ($\omega>0$), and the
fundamental p-mode has the largest $\omega$.

\item Narrow rings have a trivial slow g-mode, $e_K(r)=$constant. We have
examined a wide variety of broad rings and disks, but have found none that
support g-modes when they are isolated. In the presence of an external
non-Keplerian potential that supplies a fraction $1-\lambda$ of the total
precession, a variety of disks appear to support g-modes if and only if
$\lambda\lesssim 0.5$--0.6.

\item The p-modes are modulated at the characteristic spatial frequency
$k_0=\beta/r$. Thus long-wavelength p-modes occur only in disks with
substantial softening (cf. Figures \ref{fig:shapp}, \ref{fig:gauss}, and
\ref{fig:sellmode}). 

\item In general disks support a finite number of discrete normal modes and a
continuum of singular modes. In some cases, such as the
Jacobs-Sellwood ring with $\beta>0.229$, there are no discrete normal modes.

\end{enumerate}

Of course, these models have many shortcomings: they are linearized; they use
softened gravity as a crude proxy for velocity dispersion or pressure; and the
softening law (eq. \ref{eq:ooppsoft}) was chosen for mathematical convenience
and does not correspond to a simple interparticle force law. The models also
do not address the important questions of how rapidly slow modes damp
(although the N-body simulations of Jacobs and Sellwood 1999\nocite{js99}
suggest they can be very long-lived), or how large the disk-mass ratio $M_d/M$
can be in disks that support discrete slow modes. To decide these issues there
is no substitute for careful $N$-body simulations; however, our results
provide a guide for interpreting these simulations. This paper also does not
address the question of how slow modes are excited and whether long-lived slow
modes actually play a significant role in the structure and evolution of a
variety of astrophysical disks.

\acknowledgements

I thank Kathryn Johnston for a number of stimulating early discussions of this
problem, and Jerry Sellwood and Alar Toomre for thoughtful comments. This
research was motivated and guided by the simulations of eccentric disks
carried out by Vince Jacobs and Jerry Sellwood, and I thank them for sharing
their results freely. This work was supported in part by NSF grant AST-9900316
and NASA Grant NAG5-7310.

\end{document}